\documentclass[manuscript]{acmart}
\renewcommand\footnotetextcopyrightpermission[1]{}

\AtBeginDocument{%
  \providecommand\BibTeX{{%
    \normalfont B\kern-0.5em{\scshape i\kern-0.25em b}\kern-0.8em\TeX}}}

\setcopyright{rightsretained}

\newcommand{\sysname}{PAIR}

\usepackage{multirow}
\usepackage{subcaption}
\usepackage{caption}
\usepackage{booktabs}
\usepackage{array}
\usepackage{pifont}

\definecolor{GoodGreen}{RGB}{0,128,0}
\usepackage[nointegrals]{wasysym}
\usepackage{tabularx}

\newcolumntype{C}{>{\centering\arraybackslash}m{.047\textwidth}}

\newcommand{\hasfull}{%
  \makebox[1em][c]{%
    \textcolor{GoodGreen}{%
      \scalebox{1.25}{\ding{51}}%
    }%
  }%
}

\newcommand{\haspart}{%
  \makebox[1em][c]{\textcolor{black}{\Circle}}%
}

\begin{document}

% \title[PAIR: Perceptual Affective Inference and Regulation]{PAIR: Perceptual Affective Inference and Regulation in a Real-Time Multimodal Conversational Agent}
\title[From Momentary Emotion Inference to Sustained Emotion Support]{From Momentary Emotion Inference to Sustained Emotion Support: Evaluating a Companion Agent in a Longitudinal Study}

% \author{Anonymous}
% \affiliation{%
%   \institution{Anonymous}}

% \renewcommand{\shortauthors}{Anonymous, et al.}

\author{Kexin Quan}
\email{kq4@illinois.edu}
\affiliation{%
  \institution{School of Information Sciences, University of Illinois Urbana-Champaign}
  \city{Champaign}
  \state{Illinois}
  \country{United States}}

\author{Zijian Ding}
\email{ding@umd.edu}
\affiliation{%
  \institution{College of Information, University of Maryland}
  \city{College Park}
  \state{Maryland}
  \country{United States}}

\author{Jiaye Yong}
\email{jiayey3@illinois.edu}
\affiliation{%
  \institution{University of Illinois Urbana-Champaign}
  \city{Champaign}
  \state{Illinois}
  \country{United States}}

\author{Qinshi Zhang}
\email{qz26a@fsu.edu}
\affiliation{%
  \institution{Department of Computer Science, Florida State University}
  \city{Tallahassee}
  \state{Florida}
  \country{United States}}

\author{Dong Wang}
\email{dwang24@illinois.edu}
\affiliation{%
  \institution{University of Illinois Urbana-Champaign}
  \city{Champaign}
  \state{Illinois}
  \country{United States}}

\author{Jessie Chin}
\email{chin5@illinois.edu}
\affiliation{%
  \institution{School of Information Sciences, University of Illinois Urbana-Champaign}
  \city{Champaign}
  \state{Illinois}
  \country{United States}}

\renewcommand{\shortauthors}{Quan et al.}

\begin{abstract}
Sustained emotional support is a long-horizon interaction task closely tied to human well-being. Recent research demonstrates generative agents' capacity for momentary emotional support, yet how these capabilities translate into sustained support over time remains unclear. To examine this challenge, we deployed \sysname{}, a theory-based emotion-regulation companion, with 19 participants for 14 days. Across 1{,}093 sessions, we paired emotion estimates with self-reports before and after guidance and analyzed logs and interviews. Estimates corresponded more closely to self-reported valence and dominance than arousal. Guided conversations were followed by higher valence and state-dependent arousal changes. Participants felt understood through contextual exploration and emotional acknowledgment, acting on guidance suited to their needs and constraints. Perceived helpfulness of guided conversation significantly increased over time. Our findings link memory updates and retained corrections to cross-session personalization, informing future emotional support tools that adapt to evolving needs, learn from prior outcomes, and preserve user control over memory.

\end{abstract}




\begin{teaserfigure}
\centering
\includegraphics[width=\textwidth]{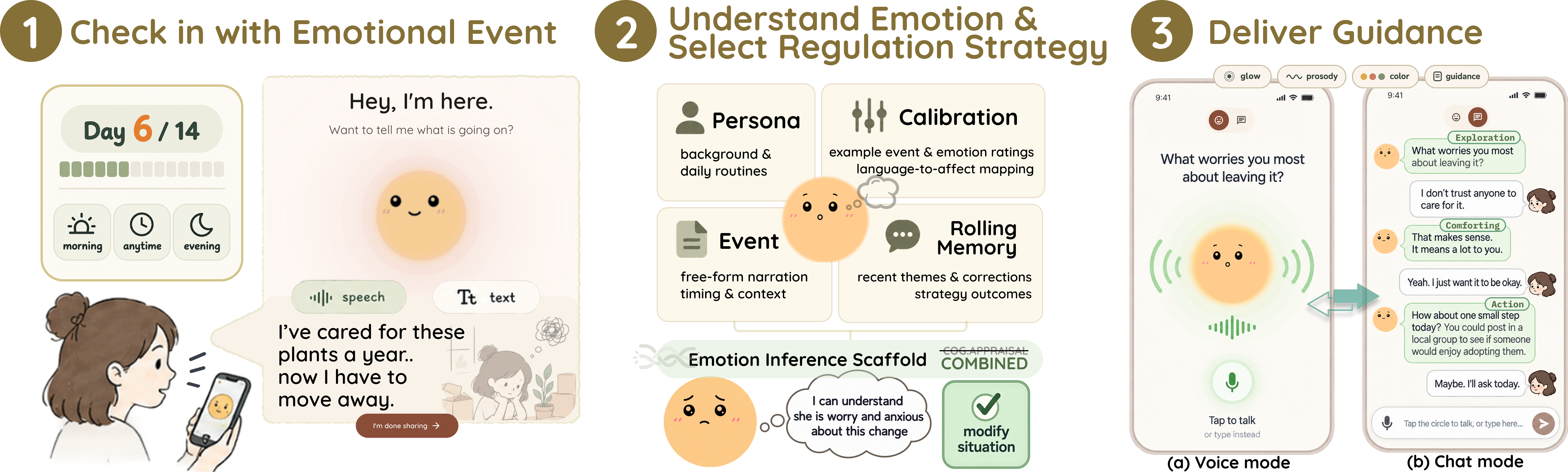}
  \caption{\sysname{} supports users across everyday emotional moments over 14 days. (1) Users can check in whenever something happens and share the experience through speech or text. (2) The companion draws on what it has learned about the user, including persona, calibration, and prior conversations, to understand how they are feeling and choose a fitting regulation strategy. (3) It then guides the user through the moment in either voice or chat, helping them move toward a more manageable next step with more support.}
  \Description{A three-stage illustration of a user's experience with \sysname{} over a 14-day deployment. A user checks in after an emotional event and shares what happened through speech or text. The companion uses information about the user and prior interactions to interpret the emotion and choose a regulation strategy. It then provides guidance through voice or chat, helping the user identify a concrete next step and feel more supported.}
  \label{fig:teaser}
\end{teaserfigure}

\maketitle

\section{Introduction}
\label{sec:intro}

Generative agents demonstrate growing capabilities in interpreting and responding to emotions within individual conversations \cite{broekens2023fine,sabour2024emobench,liu2021towards}, yet translating these capabilities into sustained support for well-being remains a challenge \cite{held2025socrates,abbas2026pitch,jorke2026bloom}. Language models perform appraisal-based emotion reasoning \cite{tak2023gpt,yongsatianchot2023investigating,kim2025modular} and generate psychologically grounded reappraisals \cite{zhan2024reappraisal}, supporting interactive tools for cognitive restructuring \cite{sharma2023cognitive,sharma2024facilitating}. For people managing recurrent stressors or interpersonal difficulties, however, emotion regulation involves ongoing efforts to manage affect across changing events and circumstances \cite{gross2015emotion,wadley2020digital,smith2022digital,slovak2023designing}. Sustained support therefore has to adapt to evolving needs and use relevant context from earlier interactions to inform subsequent responses \cite{nahumshani2018just,jo2024ltm}. Recent deployments reveal difficulties in maintaining this continuity, including repetitive questioning during cognitive reappraisal and unmet expectations of recall in daily planning and reflection \cite{held2025socrates,abbas2026pitch}.

Addressing these challenges requires an agent to translate its \textbf{understanding} of the user's emotions into \textbf{guidance} that remains appropriate as circumstances and needs change. A recurring concern may initially call for reassurance and later require practical assistance \cite{liu2021towards,cheng2022multiesc}. Deciding what to offer next therefore depends on both the user's current circumstances and their responses to earlier guidance. Continued interaction provides opportunities to learn about these developments as users revisit concerns, explain what helped, or describe difficulties acting on suggestions \cite{abbas2026pitch,skjuve2023disclosure,zheng2025customizing}. Maintaining \textbf{continuity} requires the agent to incorporate this feedback into its understanding of the user's situation and adjust subsequent guidance accordingly. Cross-session memory and personalization provide mechanisms for retaining and applying this evolving context \cite{liu2024compeer,jo2024ltm,spitale2025vita}, allowing later support to build on earlier encounters while remaining responsive to current needs.
We therefore ask: \textbf{How can generative agents translate momentary emotion understanding and regulation capabilities into sustained emotional support for individuals over time?} We examine this question through three research questions:
\begin{enumerate}
    \item \textbf{RQ1 (Emotion Understanding):} How accurately can text-based appraisal-guided inference estimate users' valence, arousal, and dominance from everyday event narratives?
    \item \textbf{RQ2 (Regulation Guidance):} How do momentary emotional outcomes and guidance uptake vary with initial emotional state, regulation strategy, and guidance fit?
    \item \textbf{RQ3 (Cross-Session Continuity):} How do conversational familiarity and accumulated personal context shape personalization and engagement over repeated interactions?
\end{enumerate}

To investigate these questions, we developed \sysname{} (Perceptual Affective Inference and Regulation), an emotional support companion combining theory-based emotion inference, state-conditioned regulation guidance, and cross-session memory and adaptation. The companion interprets a narrated event, guides a brief supportive conversation, and carries personal context and feedback into subsequent encounters. A formative human evaluation informed the selection of its inference prompts. We then conducted a 14-day diary-style field deployment in which 19 participants completed 1{,}093 sessions. We compared emotion estimates with contemporaneous self-reports before and after guidance, grounding inference evaluation in the experience of the person receiving support \cite{park2020kemocon}. Participants evaluated guidance through immediate feedback and reports of subsequent uptake. Interaction logs, questionnaires, and interviews captured how participants’ disclosure practices, support expectations, and perceived personalization evolved over sustained interactions.

The findings connect momentary performance with the interactional conditions of sustained support. Emotion estimates corresponded more closely to self-reported valence and dominance than arousal. Perceived understanding had little association with numerical error; participants described feeling understood through contextual exploration and emotional acknowledgment. Guided conversations were followed by state-dependent emotional changes, including higher valence and lower arousal in anxious sessions, with guidance uptake shaped by practical feasibility and immediate support needs. Across the deployment, session frequency showed no reliable trend, while user inputs became shorter and helpfulness ratings increased. Remembered experiences and regular dialogue supported personalization, while forgotten corrections and repetitive advice exposed difficulties in adapting support to changing circumstances.
This work contributes:
\begin{itemize}
    \item \textbf{Emotional Support Companion:} We present \sysname{}, an emotional support companion that routes everyday event narratives to appraisal-based reasoning scaffolds and selects regulation strategies using emotional intensity, controllability, and event timing. Its cross-session memory incorporates recurring concerns, user corrections and guidance outcomes to inform subsequent emotion inference and supportive conversations.
    \item \textbf{Longitudinal Field Evidence:} We provide mixed-methods evidence from a 14-day deployment with 19 participants across 1{,}093 sessions, pairing emotion estimates with users' self-reports and tracing guidance uptake and repeated use through feedback, logs, and interviews. The analysis identifies how conversational acknowledgment supports perceived understanding, practical constraints shape guidance uptake, and remembered experiences and forgotten corrections shape personalization.
    \item \textbf{Design for Sustained Support:} We derive design implications for aligning dialogue with evolving support goals and adapting guidance to prior attempts and outcomes. We further specify how inspectable, revisable memory and user control over retention and reuse can support sustained personalization.
\end{itemize}

\section{Related Work}
\label{sec:related}

\subsection{Emotional Support Systems and Companion Agents}
\label{sec:rw-companions}

Generative agents provide emotional support through opportunities for personal disclosure, reflective dialogue, coping assistance, and companionship \cite{sanches2019hci,kim2024mindfuldiary,liu2024compeer}. Conversational journaling and adaptive questioning help users articulate experiences and feelings, while behavioral sensing situates reflection within daily activities and routines \cite{kim2024mindfuldiary,seo2024chacha,song2025exploreself,nepal2024mindscape}. These interactions also shape how experiences are expressed, as question design and Artificial Intelligence (AI)-assisted writing influence users' accounts \cite{wei2024leveraging,kim2024diarymate}. Research on disclosure and empathy further examines how sharing personal experiences and receiving responsive dialogue contribute to perceived support \cite{ho2018psychological,meng2021emotional,cuadra2024illusion}. Coping assistance extends this support by helping users identify ways to respond to emotional difficulties. Human-Computer Interaction (HCI) research connects this support to everyday regulation practices and intervention design \cite{slovak2023designing,smith2022digital}. Chatbots offer psychotherapy (such as cognitive behavioral therapy), mindfulness exercises, and cognitive reappraisal, with evaluations examining acceptability, therapeutic alliance, and symptom management \cite{fitzpatrick2017woebot,inkster2018wysa,sharma2024facilitating,held2025socrates,heinz2025therabot}. Positive-psychology exercises and brief coping activities provide additional forms of support \cite{spitale2025vita,jeong2023jibo,paredes2014poptherapy}. Just-in-time adaptive intervention (JITAI) frameworks further advance inferences of user conditions for delivering timely proactive assistance \cite{nahumshani2018just}.

Personalized and expressive interaction shapes how emotional support is offered and experienced within a conversation. Peer-support agents and configurable personas accommodate users' preferred styles of support \cite{liu2024compeer,zheng2025customizing}. Voice and embodied expression provide additional ways to convey care and responsiveness \cite{zhu2022effects,loveys2021effects}, while prior study, VITA, adapts robotic coaching using facial valence and speech duration \cite{spitale2025vita}. These designs connect supportive content with how it is communicated to the user. As shown in Table~\ref{tab:rw-systems}, the selected deployments span conversational journaling, coping assistance, and mental well-being coaching, combining guided dialogue, proactive contact, and adaptation across repeated encounters. These systems inform our integration of emotional understanding, coping guidance, and expressive interaction in \sysname{}.

\begin{table*}[t]
\caption{Comparison of selected longitudinal language-model-based emotional support systems.
Icons: {\protect\hasfull} yes, {\protect\haspart} partial, blank = not reported.
\textbf{Emo.} = emotion regulation as the target;
\textbf{State.} = explicit state representation;
\textbf{Supp.} = multiple support strategies;
\textbf{Cond.} = state-conditioned strategy selection;
\textbf{Mem.} = cross-session user memory;
\textbf{Adapt.} = memory updated from subsequent user feedback;
\textbf{Schedule} = deployment duration and interaction frequency.}
\label{tab:rw-systems}
\centering
\footnotesize
\setlength{\tabcolsep}{3.5pt}
\renewcommand{\arraystretch}{1.15}
\begin{tabular}{@{}>{\raggedright\arraybackslash}p{.175\textwidth}*{6}{>{\centering\arraybackslash}p{.047\textwidth}}>{\centering\arraybackslash}p{.10\textwidth}>{\raggedright\arraybackslash}p{.28\textwidth}@{}}
\toprule
& \multicolumn{2}{c}{\textbf{Inference}} & \multicolumn{2}{c}{\textbf{Guidance}} & \multicolumn{2}{c}{\textbf{Continuity}} & & \\
\cmidrule(lr){2-3}\cmidrule(lr){4-5}\cmidrule(lr){6-7}
\textbf{System} & \textbf{Emo.} & \textbf{State.} & \textbf{Supp.} & \textbf{Cond.} & \textbf{Mem.} & \textbf{Adapt.} & \textbf{Schedule} & \textbf{What the system does} \\
\midrule
% MindfulDiary supports emotional expression through journaling; emotion summaries
% provide a partial state representation. The paper describes within-session
% context and clinician-facing records, not cross-session chatbot memory.
MindfulDiary (CHI'24) \cite{kim2024mindfuldiary}
& \haspart & \haspart & & & & & 28 d, on demand &
Conversational journaling with emotion summaries for patients and clinicians. \\
\addlinespace
Socrates 2.0 (JMIR MH'25) \cite{held2025socrates}
& \hasfull & & & & & & 28 d, on demand &
Socratic cognitive reappraisal for depression and anxiety. \\
\addlinespace
% Therabot targets clinical symptoms and provides empathy, validation, and
% targeted interventions. Conversation history provides partial memory support;
% explicit emotion-state routing and feedback-updated memory are not reported.
Therabot (NEJM AI'25) \cite{heinz2025therabot}
& \haspart & & \haspart & & \haspart & & 4 wk daily; 4 wk on demand &
Generative mental health support using expert-trained dialogue and conversation history. \\
\addlinespace
ComPeer (UIST'24) \cite{liu2024compeer}
& \hasfull & & & & \hasfull & \haspart & 7\,+\,7 d, proactive &
Proactive peer support from a persona and conversation history. \\
\addlinespace
VITA (THRI'25) \cite{spitale2025vita}
& \hasfull & \hasfull & \haspart & \hasfull & \haspart & \hasfull & 28 d $\times$ 1/week &
Robotic well-being coaching through positive-psychology exercises. \\
\midrule
\textbf{\sysname{} (our work)}
& \hasfull & \hasfull & \hasfull & \hasfull & \hasfull & \hasfull & 14 d, daily + events &
Appraisal-guided emotion regulation across everyday events. \\
\bottomrule
\end{tabular}
\end{table*}

\subsection{Single-Session Capabilities and Limits in AI Emotional Support}
\label{sec:rw-capabilities}

Research on AI emotional support has largely been conducted in a single-session setting. These studies examine how language models infer emotions from dialogue corpora and generate supportive responses. Zero-shot prompting yields dimensional affect ratings and appraisal-based inferences that align with human appraisals and emotion labels, with prompt structure shaping recognition performance \cite{broekens2023fine,tak2023gpt,yongsatianchot2023investigating,kim2025modular}. Appraisal theory also guides the generation of cognitive reappraisals and organizes dialogue into exploration, comforting, and action, producing supportive replies that experts evaluate favorably \cite{zhan2024reappraisal,hu2024aptness,zhang2024escot,liu2021towards,cheng2022multiesc}. Evaluations of these single exchanges judge generated support favorably against human comparison points on empathy, compassion, and emotional awareness \cite{sharma2023cognitive,ovsyannikova2025third,elyoseph2023chatgpt}. 

However, single-session evaluation cannot show whether support holds up over time. Emotional support research models limited-turn or single-session dialogue with little representation of long-term user trajectories, while support in practice spans days or weeks and requires tracking evolving states and integrating implicit, fragmented disclosures across sessions \cite{chen2026esmemeval}. Outcome evidence carries the same boundary. A meta-analysis of 32 randomized controlled trials found significant short-term effects on depressive and anxiety symptoms but no significant long-term effects, with most trials reporting no follow-up beyond 8 weeks \cite{he2023agents}. The same analysis also found personalization, empathic response, and longer interaction associated with larger effects \cite{he2023agents}. Another meta-analysis of AI-based conversational agents likewise reports symptom reduction without significant improvement in overall psychological well-being \cite{li2023metaanalysis}. Overall, this evidence supports short-term AI emotional support capabilities but stops short of showing whether single-session performance carries over as circumstances, needs, and prior guidance accumulate.

\subsection{Cross-Session Continuity in AI Support Systems}
\label{sec:rw-long-horizon}

Sustained AI support builds on repeated interactions around daily routines, personal goals, and shared activities. Field deployments have examined robots as social partners in schools and facilitators of parent--child reading at home \cite{kanda2004field,chen2025catalysts}. Conversational companions and coaches extend these roles to daily planning, reflection, and physical activity \cite{abbas2026pitch,jorke2026bloom}. Within these recurring activities, personal memory and proactive contact help establish continuity across encounters. In public health check-ins, CareCall's long-term memory supported familiarity and greater health disclosure \cite{jo2024ltm}. PITCH's morning planning and evening reflection conversations elicited acceptance, negotiation, and resistance to the agent's suggestions, revealing how users responded to its proactive role \cite{abbas2026pitch}. Maintaining this continuity requires retaining and updating personal information from earlier encounters. Conversational-memory benchmarks document difficulties in retrieving relevant information, reasoning across sessions, updating knowledge, and tracking changing preferences \cite{maharana2024locomo,wu2025longmemeval,personamem2025}. These studies connect continued interaction to how a companion remembers the user and participates in their routines.

Longitudinal studies show that companion relationships follow varied trajectories as users' expectations, disclosures, and reasons for returning evolve. Research on social chatbots documents growing closeness alongside disruptions in trust, immersion, and relationship development associated with inconsistent responses and unexpected system behavior \cite{croes2021mitsuku,skjuve2022longitudinal,lopeztorres2023}. A twelve-week disclosure study found narrowing conversational breadth and varied trajectories of conversational depth \cite{skjuve2023disclosure}. Users also bring different expectations to these relationships, as research with autistic and non-autistic adults shows for empathy and conversational depth \cite{xygkou2024social}. Repeated use also exposed repetitive questioning and limited conversational depth in Socrates 2.0 \cite{held2025socrates}. This work motivates examining how changing expectations affect the fit of subsequent emotional support and whether memory and user feedback help guidance remain relevant.

\section{Technical Evaluation of AI Emotional Understanding}
\label{sec:technical-evaluation}

Recent work has demonstrated strong momentary emotional understanding in generative agents for emotional support~\cite{broekens2023fine, tak2023gpt, elyoseph2023chatgpt}. Such understanding is fundamental to both effective regulation guidance and sustained support. Because emotion is person-centered, evaluating whether an interpretation reflects a person's actual experience requires human judgment~\cite{park2020kemocon}. We therefore conducted a formative human evaluation of four prompting approaches on a state-of-the-art model \cite{anthropic2026claude}, asking raters with first-hand knowledge of the events to judge the plausibility and fidelity of each interpretation. The prompts translate cognitive-appraisal and neurocognitive theories into increasingly structured reasoning scaffolds, allowing us to test how theory-grounded prompting improves emotion inference and to select the designs deployed in \sysname{}.

\subsection{Experimental setup}
\label{sec:prompteval}

\subsubsection{Four Theory-based Prompt Designs}
\label{sec:promptdesign}

Building on evidence that explicit appraisal guidance can structure LLM emotion reasoning and cognitive reappraisal~\cite{yongsatianchot2023investigating, zhan2024reappraisal}, we designed four prompts with incremental reasoning scaffolding (Figure~\ref{fig:prompteval}a). The full text of each prompt is provided in the supplementary materials:

\begin{enumerate}
    \item \textbf{Baseline}: adapted from modular prompt designs for LLM-based emotion recognition \cite{kim2025modular}, asks the model to infer the emotional state directly without prescribed reasoning;
    \item \textbf{Cognitive Appraisal Inference (Cog)}: structures reasoning as a five-stage cognitive-appraisal chain following Lazarus and Scherer \cite{lazarus1991emotion, scherer2001appraisal}: situational meaning, goals and expectations, expectation violation, appraisal dimensions (goal relevance and congruence, coping potential, agency, novelty), and integration;
    \item \textbf{Neurocognitive Emotion Inference (Neuro)}: draws on broad neuroscience accounts of emotion processing, structuring inference around salience detection, memory retrieval, and prefrontal appraisal and regulation \cite{ledoux2000emotion, mcgaugh2004amygdala, ochsner2005cognitive};
    \item \textbf{Combined}: couples bottom-up neurocognitive processing with top-down appraisal in a single emotion-inference scaffold \cite{pessoa2008relationship}. The four prompts differ only in their prescribed reasoning and share an identical output format. Each returns an initial emotion estimate with discrete labels, intensity, and pleasure (valence)-arousal-dominance (PAD) values, a guidance message, and a prediction of emotion after the proposed guidance.
\end{enumerate}

\begin{figure}[t]
\centering
\includegraphics[width=\textwidth]{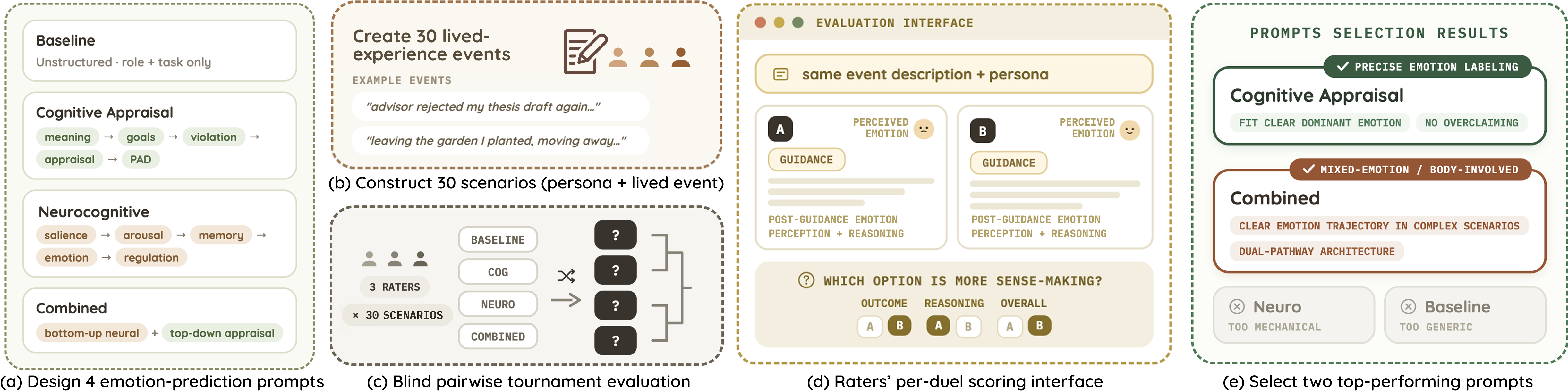}
  \caption{Prompt evaluation pipeline. (a) Four emotion-inference prompts with incremental reasoning scaffolding are compared on (b) 30 scenarios pairing personas with lived-experience events, through (c, d) blind pairwise bracket tournaments by three human raters, (e) yielding two top performers with complementary strengths, both deployed in \sysname.}
  \Description{A five-panel diagram showing four prompt designs, the construction of 30 persona and event scenarios, a blind pairwise prompt-evaluation bracket, the raters' scoring interface with three rating dimensions, and the two winning prompts selected for deployment.}
  \label{fig:prompteval}
\end{figure}
\subsubsection{Procedure}

\label{sec:evalprocedure}

We constructed 30 scenarios for the prompt evaluation (Figure~\ref{fig:prompteval}b). Three researchers each contributed ten real-life events from their own lives or those of people close to them. Each event was paired with an extended narrator persona to form one scenario. The same researchers served as raters, ensuring that each scenario had at least one rater with first-hand or close-person knowledge of the contributed experience (Figure~\ref{fig:prompteval}c). All four prompts used Claude Opus 4.8~\cite{anthropic2026claude} and were evaluated in a blinded bracket tournament across all scenarios. Holding the model constant isolated differences in reasoning scaffolds. In each comparison, raters viewed the persona, event, and two anonymized outputs side by side (Figure~\ref{fig:prompteval}d). The underlying reasoning was available on demand. Guidance was shown only as context for evaluating the predicted post-guidance emotion and was not itself rated. Each rater completed the tournament for every scenario, yielding 90 scenario-rater rankings (30 scenarios $\times$ 3 raters) and 1{,}080 dimension-level pairwise judgments across the three rating dimensions.

\subsubsection{Measures and Analysis}

Each comparison used three forced-choice measures. \textit{Outcome plausibility} assessed whether the predicted post-guidance emotion was realistic for the person and event. \textit{Reasoning fidelity} assessed whether the explanation reflected the person's actual thoughts and circumstances. \textit{Overall preference}, our primary measure, integrated both judgments and required a written justification. Because results were consistent across the three measures, we focus on overall preference. The tournament produced a complete ranking of the four prompts for each rater and scenario. For the comparison between Cog and Combined, we identified which prompt each rater ranked higher and used majority vote across the three raters to determine the scenario-level winner. We then coded all 30 events on four binary features using only the event text: single versus multiple events, single versus mixed emotions, goal clarity, and persona dependence. Coding was completed before examining prompt preferences. We cross-tabulated each feature with the Cog-versus-Combined winner and report counts and proportions. Finally, we examined all scenarios covering clear Cog wins, clear Combined wins, boundary cases, and rater disagreement. We compared the paired outputs with raters' written justifications to identify the reasoning patterns behind each preference (Figure~\ref{fig:prompteval}).

\begin{figure}[t]
\centering
\includegraphics[width=0.9\textwidth]{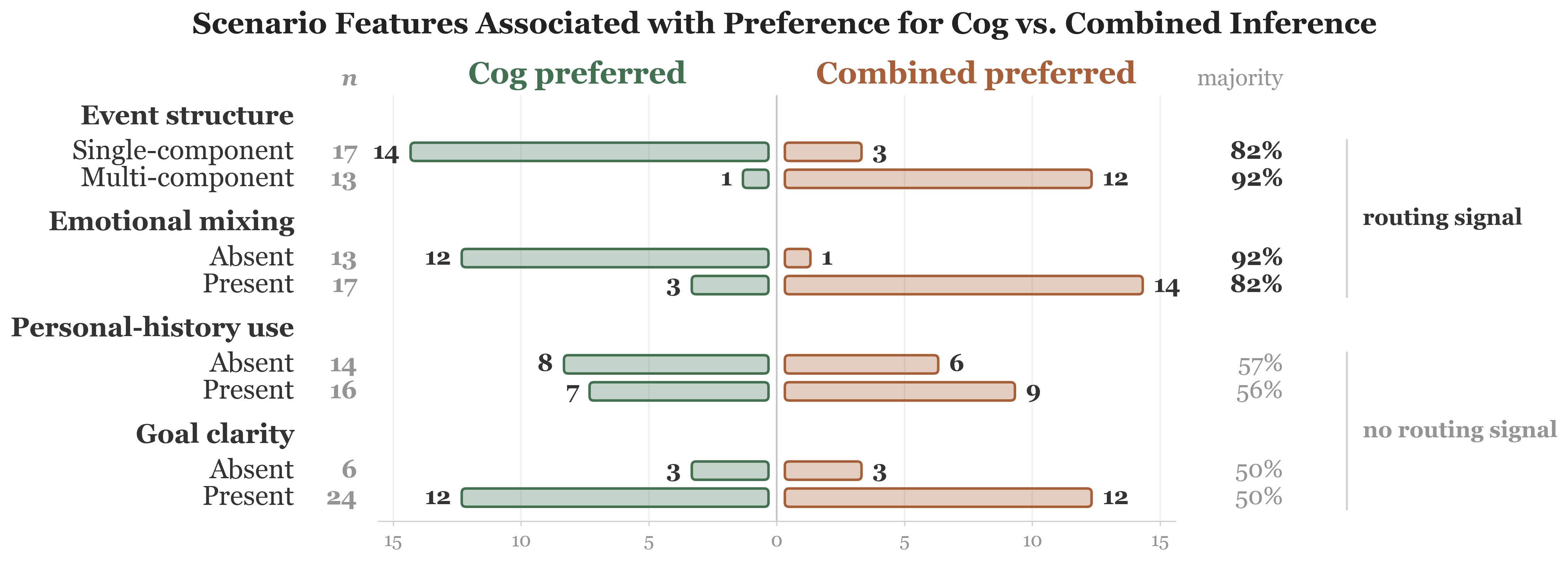}
\caption{Event structure and emotional mixing were associated with prompt preference in the 30-scenario formative set, while preferences were more evenly divided for personal-history use and goal clarity. Event structure and emotional mixing largely overlapped, and the deployed router uses event structure alone. Bars give the number of scenarios in which each prompt was preferred by majority vote based on the raters' rankings; the right column reports the winning share.}
\Description{A butterfly chart with eight rows grouped into four scenario
features. Scenarios won by Cog extend left in green, those won by Combined
extend right in orange. For event structure and emotional mixing the long bar
switches sides between the feature's two levels; for personal-history use and
goal clarity both levels show short, near-equal bars on either side.}
\label{fig:promptcoding}
\end{figure}

\subsection{Results: Cognitive Appraisal Inference for Single-Component Events and Combined Inference for Multi-Component Events}
\label{sec:evalfindings}

\subsubsection{Appraisal-grounded prompts outperformed the alternatives} Both appraisal-informed prompts ranked above Baseline and Neuro overall (full ranking distribution in Appendix~\ref{app:prompteval}). Cog achieved the best mean rank and the most consistent performance, while Combined received the second-most first-place rankings. Both won most direct comparisons with Baseline, which was most frequently ranked last. Cog also won most comparisons with Neuro. Raters often characterized Neuro's analyses as mechanistic and disconnected from how people described their emotional experiences. Explicit appraisal structure therefore produced stronger emotion interpretations than direct estimation or neuroscience-based emotion reasoning alone.

\subsubsection{Cog and Combined Scaffolds performed best on different event structures} Cog and Combined were equally preferred in scenario-level comparisons derived from the raters' rankings (15:15 across 30 scenarios), reflecting complementary strengths rather than equivalent performance across scenarios (Table~\ref{tab:promptprofiles}). To examine this pattern, two researchers, assisted by Claude Opus 4.8, coded each event using only its text and without reference to the ranking outcomes, then reviewed and finalized all assignments. Cog was preferred for nearly all events with a single trigger, a clear causal chain, and one dominant emotional response. Combined was preferred for nearly all events involving multiple interacting events or emotional components. Event structure matched the majority-preferred prompt in 26 of the 30 formative scenarios (87\%; Figure~\ref{fig:promptcoding}). The four mismatches occurred in scenarios where raters disagreed.

\subsubsection{Scaffold complexity should match event structure} Raters' overall judgments aligned with the reasoning dimension in all 30 scenarios, compared with 27 scenarios for the predicted outcome. Prompt preference therefore followed reasoning quality more consistently than the predicted post-guidance emotion. The prompts' failure modes further explained the importance of structural fit (Table~\ref{tab:promptprofiles}). For multi-component events, Cog often omitted relevant bodily influences and compressed layered emotions into a single causal account. For simpler events, Combined introduced unsupported connections to the persona or prior experiences. Raters penalized this added complexity when the event did not require integration but accepted it when the event contained genuinely interacting components. We therefore retained both prompts and routed each event to the scaffold that matched its structure.

Based on these findings, \sysname{} uses a lightweight router to select Cog for single-component events and Combined for events with multiple interacting components (Figure~\ref{fig:prompteval}e).

\begin{table}[t]
\caption{Complementary strengths and failure modes of the two selected prompts.}
\label{tab:promptprofiles}
\small
\begin{tabular}{p{0.15\linewidth}p{0.38\linewidth}p{0.38\linewidth}}
\toprule
& \textbf{Cog (appraisal-based)} & \textbf{Combined (integrative)} \\
\midrule
Best suited for
& Single-component events (14/17): discrete incidents, isolated conflicts, and single anticipatory states
& Multi-component events (12/13): stacked stressors, chained events, body--mind interactions, and multi-stage experiences \\
\addlinespace
Primary strength
& Focused, person-centered reasoning. Follows one clear causal chain.
& Preserves multiple emotional and causal components. Captures bodily and layered experiences. \\
\addlinespace
Failure mode when mismatched
& Omits relevant components, especially bodily signals and layered emotions.
& Introduces unsupported links to the persona or past experience. \\
\bottomrule
\end{tabular}
\end{table}

\section{System Design}
\label{sec:system}

\subsection{Design Goals}
\label{sec:designgoals}

Supporting emotion regulation over a long horizon introduces requirements beyond responding to a single event. As users' emotions and situations change over time, fixed responses may become less appropriate. A long-horizon companion must first understand the user in the moment and offer suitable support. It must also carry knowledge forward so that this support can adapt over time. Both capabilities depend on sustained engagement. These considerations motivate three design goals for \sysname{}.

\textbf{(1) DG1: Provide momentary emotion understanding and support.} Following the formative evaluation in Section~\ref{sec:technical-evaluation}, \sysname{} retains both Cog and Combined and routes each narration to the scaffold that matches its event structure. The selected scaffold infers a pleasure (valence)--arousal--dominance (PAD) estimate from the user's account~\cite{mehrabian1996pleasure}. The system then selects regulation guidance based on emotional intensity, controllability, and event phase, following Gross's process model~\cite{gross2015emotion,gross1998antecedent}. This pipeline provides the momentary understanding and guidance needed for support across sessions.

\textbf{(2) DG2: Support continuity through memory and personalization.} Long-horizon support should build on what the system has learned about a user instead of treating each interaction as a fresh encounter. Prior systems demonstrate the value of incorporating personal context into ongoing emotional support and adapting the interaction to individual needs \cite{nepal2024mindscape, liu2024compeer, zheng2025customizing}. \sysname{} therefore establishes an initial user model through an onboarding persona and personal calibration, carries relevant information across sessions through memory, and updates its understanding and guidance in response to the user's subsequent accounts, corrections, and outcomes. This continuity allows momentary support to become progressively grounded in the user's experiences and preferences.

\textbf{(3) DG3: Sustain engagement across sessions.} Continued interaction is a functional requirement for long-horizon support: users must be willing to return and share new information for the system's understanding and personalization to improve. In-situ affective data depend on users' willingness to provide it, and diary-style protocols degrade when reporting feels like work \cite{bolger2003diary}. Prior emotion technologies have therefore emphasized lightweight, in-the-moment capture \cite{sanches2019hci, nepal2024mindscape, kim2024mindfuldiary}, while visual and auditory cues can make affective feedback more immediate and engaging \cite{costa2016emotioncheck, choi2020ambienbeat, yu2018delight, spiridon2017effects}. \sysname{} supports speech and typed input, a visual avatar, and coordinated multimodal feedback to lower the effort of everyday interaction and present guidance in an expressive, approachable form. These features create the conditions for sustained use, enabling the system to gather the cross-session information on which sustained understanding and adaptation depend.

\begin{figure}[t]
\centering
\includegraphics[width=\textwidth]{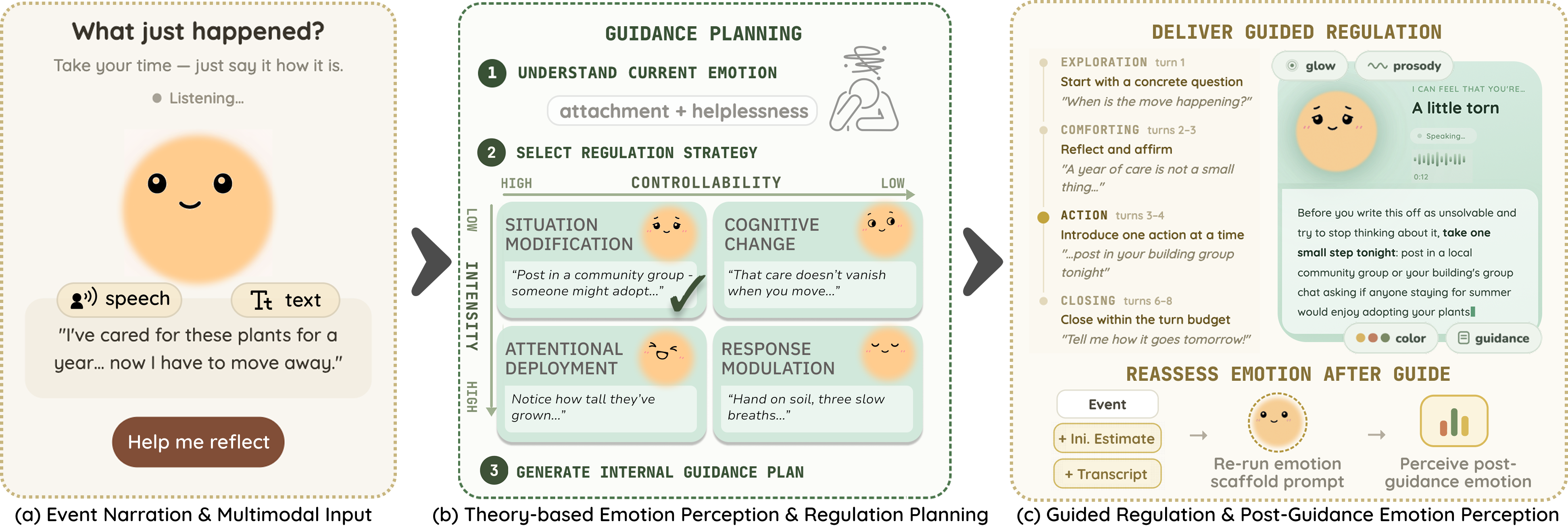}
\caption{User interaction flow and guidance logic in \sysname{}.
(a) The user recounts an everyday event through speech or text.
(b) The system perceives the user's current emotional state using one of two theory-grounded reasoning scaffolds, selects a regulation strategy based on the event's intensity and controllability, and generates an internal guidance plan.
(c) The plan is delivered through a brief guided conversation with coordinated multimodal cues. After guidance, the system produces a post-guidance estimate from the original event, the initial estimate, and the conversation transcript.}
\Description{A three-panel diagram showing the user interaction flow and core guidance logic in \sysname{}. Panel (a) shows a user describing an everyday event through speech or text. Panel (b) shows the system perceiving the user's current emotional state, selecting among four regulation-strategy families based on intensity and controllability, and generating an internal guidance plan. Panel (c) shows the plan delivered through a short guided conversation with multimodal cues, followed by a post-guidance estimate produced from the original event, the initial emotion estimate, and the conversation transcript.}
\label{fig:pipeline}
\end{figure}

\subsection{System Implementation}

\label{sec:implementation}

We implemented \sysname{} as a responsive web application for desktop and mobile browsers. Claude Opus 4.8~\cite{anthropic2026claude} handles emotion inference, guided conversation, and post-guidance estimation, while GPT-5-mini~\cite{openai2026gpt5mini} handles routing, memory summarization, daily tips, and avatar expression selection. Voice input is transcribed in the browser using the Web Speech API, with no raw audio stored. GPT-4o-mini-TTS~\cite{openai2026gpt4ominitts} is used to generate spoken responses. Figure~\ref{fig:orchestration} in Appendix~\ref{app:orchestration} summarizes the within-session orchestration and cross-session memory updates. The following sections describe how the architecture implements each design goal.

\subsubsection{Momentary Emotion Understanding and Support (DG1)}
\label{sec:momentary-support}

\textbf{Theory-grounded emotion understanding.}

Following Section~\ref{sec:prompteval}, a lightweight router assigns single-component narrations to Cog and multi-component narrations to Combined. The selected scaffold infers the user's emotional state as a PAD estimate and corresponding 9-point Self-Assessment Manikin (SAM) scores for valence, arousal, and dominance~\cite{bradley1994measuring}, together with emotion labels, intensity, controllability, and temporal phase. The inference also conditions on the user's persona, calibration example, and cross-session memory described in DG2.
\textbf{Context-sensitive regulation guidance.}
Using this inferred state, the system selects among four regulation-strategy families from Gross's process model: situation modification, attentional deployment, cognitive change, and response modulation~\cite{gross2015emotion}. Strategy selection is primarily conditioned on emotional intensity and controllability, with temporal phase, interaction timing, and personal context providing additional constraints. The selected strategy is then operationalized as a concise internal guidance plan containing its contextual rationale and an actionable next step. A conversational agent delivers this guidance through an Exploration--Comforting--Action (ECA) sequence adapted from emotional support conversation frameworks~\cite{liu2021towards}. The interaction first elicits relevant situational context, then acknowledges the user's emotional experience before introducing regulation guidance. The agent adapts its responses as new information emerges and moves toward action with a target of approximately ten turns.
\textbf{Reassessing emotion after guidance.}
After the conversation, the appraisal scaffold re-estimates the user's PAD and SAM state using the original event, initial estimate, and complete conversation. The scaffold assesses how the interaction changed the user's emotional state while conditioning on their persona and context, without presuming improvement. This post-guidance estimate enables paired analysis of emotional state before and after the conversation in our evaluation (Section~\ref{sec:study}).

% 3.3.2 DG2: onboarding persona + personal calibration + cross-session memory + correction and outcome-driven adaptation
\subsubsection{Continuity through Memory and Personalization (DG2)}
\label{sec:memory-personalization}

\textbf{Onboarding persona and personal calibration.}
Two onboarding components establish personal context for subsequent emotion inference and guidance. First, a seven-question persona interview captures the user's self-description, daily life, coping habits, sources of support, current concerns, and emotional tendencies. Users respond by voice or text, and a lightweight model converts their responses into a structured persona summary, a brief for the conversational agent, and a personalized speaking style. Second, a calibration exercise anchors emotion inference to the user's self-reported affect. The user narrates a recent emotional event and rates its valence, arousal, and dominance on the 1--9 SAM scales. This paired example provides an individualized reference between the user's language and affect ratings and is included as context in subsequent emotion estimates.
\textbf{Cross-session memory and adaptation.}
A rolling memory carries relevant information from prior sessions into subsequent emotion inference and guidance. After each evening session, a lightweight model summarizes the day into a structured block of up to 150 words, capturing recurring concerns, systematic differences between model estimates and self-reports, user corrections, and outcomes of prior regulation strategies. This memory block is included in later appraisal and conversation calls, so that previous estimation errors can inform later calibration, corrections can revise the system's interpretation, and prior outcomes can shape future guidance. The resulting feedback loop supports adaptation of both emotion understanding and regulation guidance over time.

\subsubsection{Sustained Engagement through Low-Burden and Multimodal Interaction (DG3)}
\label{sec:sustained-engagement}

\textbf{Low-burden daily interaction.}
\sysname{} supports repeated use through three type of daily touchpoints. Morning check-ins capture the user's starting state, anytime logs allow users to record emotionally significant events as they occur, and evening reflections revisit an event from the day. Each interaction begins with free-form narration through text or speech. Spoken input is transcribed in real time and remains editable before submission.
\textbf{Semantically coordinated multimodal presentation.}
\sysname{} represents the inferred emotional state and regulation target through coordinated color, pulse, speech prosody, and avatar expression~\cite{costa2016emotioncheck,yu2018delight}. Color properties reflect valence and arousal using established color--emotion associations~\cite{jonauskaite2020universal,valdez1994effects}, while avatar pulse rhythm varies with target arousal~\cite{costa2016emotioncheck,choi2020ambienbeat}. Speech prosody adapts pace, energy, pitch, and pausing to the target state and user persona. The avatar expression also updates with the evolving conversation. These channels share the same emotional representation to provide consistent visual and auditory feedback throughout the interaction.

\section{Study Design}
\label{sec:study}

To examine how \sysname{} understands and supports users' emotions in everyday life, we conducted a 14-day diary-style field deployment \cite{bolger2003diary}, pairing each system estimate with the participant's self-report of the same moment. We describe participant recruitment and characteristics (Section~\ref{sec:participants}), the study procedure (Section~\ref{sec:procedure}), and data collection and analysis (Section~\ref{sec:evaluation}).

\subsection{Participants}
\label{sec:participants}

We recruited 24 participants in the U.S. through university mailing lists and online social media. Five withdrew during the study period, resulting in a final sample of 19 participants (13 women, 6 men; aged 18--33 years, $M = 25.1$, $SD = 5.1$). Eligible participants were adults aged 18--65 who were not currently receiving psychological or psychiatric treatment. Sixty-eight percent reported high familiarity with AI tools and 42\% had previously used one for emotional support (see full participant characteristics in Appendix~\ref{app:participants}). The study received university IRB approval. Participants received up to \$80 in Amazon e-gift cards based on completed study days and participation in the post-study session. Participants were informed that the system is not a clinical tool and received a mental-health resource handout during consent.

\subsection{Study Procedure}
\label{sec:procedure}

\begin{figure}
\centering
\includegraphics[width=1.03\textwidth]{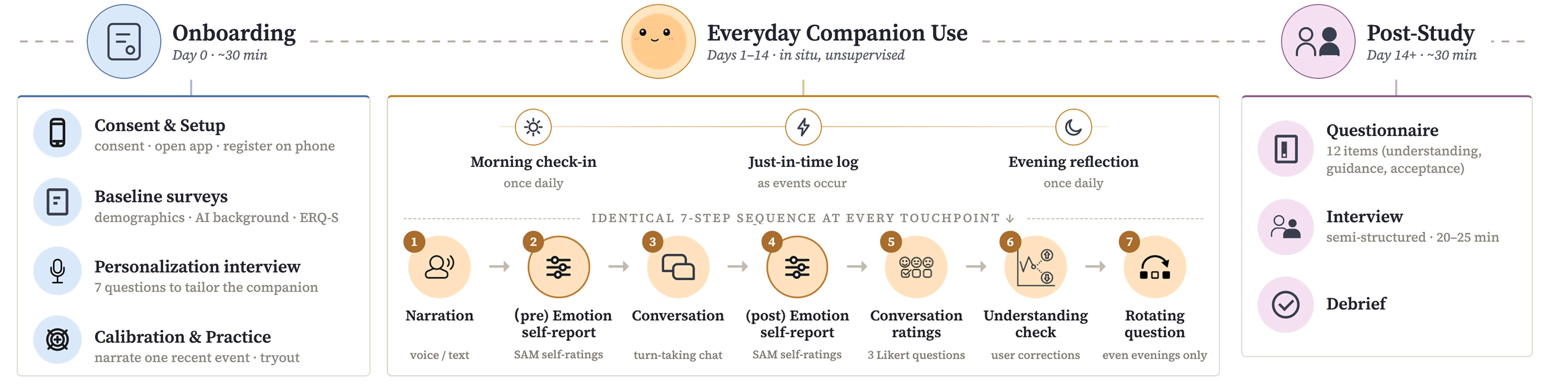}
  \caption{Study procedure. Day 0: researcher-led onboarding with consent and app setup, baseline surveys, the personalization interview, and a calibration and practice exercise. Days 1--14: three types of daily touchpoints (morning check-in, just-in-time logging, evening reflection), each pairing the system's affective estimates with SAM self-ratings before and after the guided conversation. Day 14+: post-study questionnaire and semi-structured interview.}
  \Description{A timeline diagram showing the three study phases: Day 0 onboarding, the 14-day daily companion phase with its per-session step sequence, and the post-study session.}
  \label{fig:procedure}
\end{figure}

\subsubsection{Day 0: Onboarding}
Onboarding (Figure~\ref{fig:procedure}) was conducted in a researcher-led Zoom session. After informed consent, participants created an account and completed questionnaires on demographics, AI experience, and expressive suppression using three ERQ-S items~\cite{gross2003individual} (see details in Appendix~\ref{app:participants}). Then, they completed the seven-question persona interview, covering daily life, coping habits, support sources, and current concerns (Appendix~\ref{app:persona}). Finally, participants completed the personal calibration exercise by narrating a recent emotional event and rating its valence, arousal, and dominance using SAM~\cite{bradley1994measuring}, followed by a practice just-in-time log. On these scales, valence ranged from unpleasant to pleasant, arousal from low to high activation, and dominance from low to high control. Following the circumplex model of affect, the combination of valence and arousal distinguished anxious, excited, dejected, and calm states \cite{russell1980circumplex}.

\subsubsection{Days 1--14: Everyday Companion Use}
Participants completed multiple daily touchpoints described in Section~\ref{sec:sustained-engagement}. Each interaction followed the same core procedure. Participants first narrated an event by voice or text and rated their valence, arousal, and dominance before seeing any system estimate. The estimate was generated concurrently but withheld until self-report was submitted to avoid anchoring. Participants then completed a brief guided conversation designed to take under five minutes and repeated the SAM ratings afterward. Three 7-point Likert items assessed helpfulness, feasibility, and willingness to act. Participants also reviewed the companion's emotion interpretation and provided a correction when their understanding rating was 4 or below. These corrections informed subsequent analysis and cross-session memory. Even-numbered evening sessions included a rotating probe on guidance adoption, perceived personalization, or experiences with multimodal features. The mobile-first interface supported on-the-go use, and sessions inactive for 30 minutes were automatically finalized as timed out.

\subsubsection{Day 14+: Post-Study Session}
The post-study session was conducted by a researcher over Zoom and recorded with participant consent. Participants first completed a 12-item questionnaire using 7-point Likert scales to assess perceived emotional understanding, guidance quality, and technology acceptance. The questionnaire combined study-specific items with items adapted from the Perceived Empathy of Technology Scale \cite{schmidmaier2024pets}, Session Rating Scale \cite{duncan2003session}, the cognitive-reappraisal construct \cite{gross1998antecedent}, and Technology Acceptance Model \cite{davis1989perceived}. The full questionnaire is provided in Appendix~\ref{app:questionnaire}. A 20--25-minute semi-structured interview then explored participants' overall experience, perceived accuracy of the system's emotion estimates, suggestion adoption, changes in emotional awareness, and perceptions of the multimodal features (Appendix~\ref{app:interview}).

\subsection{Data Collection and Analysis}
\label{sec:evaluation}

We adopted a mixed-methods approach combining paired system estimates and self-reports with interaction logs, session feedback, post-study questionnaires, and interviews. Quantitative analyses examined emotion-estimation accuracy, within-session affective changes, and patterns of engagement and perceived value over the deployment. Interview accounts and open-ended responses helped us examine how participants experienced the companion's understanding, guidance, and continued presence in everyday life. Appendix~\ref{app:data} summarizes the data collected.

\subsubsection{RQ1: Emotion Inference and Perceived Understanding}

We evaluated emotion-estimation accuracy against participants' paired SAM self-reports, comparing initial estimates with pre-guidance ratings and post-guidance estimates with post-guidance ratings. Valence, arousal, and dominance were assessed using MAE, Pearson's $r$, and mean signed error. We also measured lower-versus-higher state agreement using Cohen's $\kappa$, with additional thresholds examined for arousal. For initial valence, we compared the deployed appraisal scaffold with two baselines: Claude Opus 4.8, the same model without appraisal reasoning, and a lexicon-based estimator using the NRC-VAD Lexicon~\cite{mohammad2018vad}. Lexicon scores were mapped to the 1--9 SAM scale using five-fold held-out estimation, and all three estimators were evaluated against the same pre-guidance self-reports. We additionally report the NRC-VAD arousal correlation descriptively using the same held-out mapping procedure.
We also summarized perceived understanding using session-level and post-study ratings. Spearman correlations tested whether perceived understanding was associated with absolute estimation error in valence, arousal, and dominance.

\subsubsection{RQ2: Regulation Outcomes and Guidance Fit}

We measured immediate regulation as within-session change in valence, arousal, and dominance. Marginal changes were tested using exact two-sided Wilcoxon signed-rank tests on participant-level mean changes, with Holm correction across the three dimensions. We also examined joint valence--arousal trajectories across four entry states defined by a midpoint split at 5: anxious or agitated, excited or energized, dejected or flat, and calm or content. Valence change was additionally summarized for sessions starting at low ($V \leq 4$) and high ($V \geq 7$) valence. Appendix~\ref{app:marginal-affect} provides the marginal tests and visualization details.

To examine how perceived understanding and conversational alignment related to regulation outcomes, we tested the association between understanding ratings and within-session valence change using Pearson correlation and summarized mean $\Delta V$ by understanding rating. We also analyzed companion-turn stage labels, including starting stages, adjacent-stage transitions, and stage dwell time, to characterize how contextual exploration and emotional acknowledgment preceded guidance. Two researchers independently coded corrections submitted after understanding ratings of 4 or below and resolved disagreements through discussion. Codes captured emotion misreads, context or memory errors, strategy mismatches, and inability to judge the interaction. We compared each participant's mean valence change between corrected and uncorrected sessions using an exact two-sided paired Wilcoxon signed-rank test among the 12 participants contributing both session types, and descriptively compared their starting valence.

We compared participant-level mean valence changes across the four regulation-strategy families using a Friedman test. Differences in helpfulness, feasibility, and willingness to act were tested with mixed-effects likelihood-ratio tests. Model specifications and multiple-testing correction appear in Appendix~\ref{app:strategy-feedback}. Session feedback, post-study ratings, and interview accounts were used to characterize how guidance matched participants' needs and practical constraints.
Guidance adoption was assessed through evening probes asking whether participants had tried earlier suggestions. We classified 44 responses with LLM assistance~\cite{gilardi2023chatgpt} and manually reviewed each result. Categories captured successful adoption, partial or alternative enactment, already-intended actions, non-adoption, and inability to recall. We report category counts to describe uptake beyond the conversation.

\subsubsection{RQ3: Continued Use and Personalization}
To examine how engagement and perceived value changed over two weeks, we analyzed daily session frequency, turns per session, narration length, user input length per turn, and ratings of helpfulness, feasibility, and willingness to act. Mixed-effects models estimated study-day trends and Week 2 versus Week 1 contrasts (Appendix~\ref{app:trends}). Interview accounts contextualized these trends through changes in participation, expectations, and familiarity with the companion. Perceived personalization was examined through post-study ratings of personalized responses and feeling increasingly known, alongside interview accounts of cross-session memory. We examined how participants described recall of earlier interactions, updates to changing circumstances, and whether corrections carried into later conversations. Post-study ease-of-use ratings supplemented this analysis.

\subsubsection{Qualitative Analysis}
Two researchers conducted thematic analysis of the post-study interview transcripts and other rotating-question responses \cite{braun2006thematic}. They coded independently, reconciled differences, and iteratively refined the codebook. Themes were organized around research questions, with sub-themes emerging during discussion.

\section{Results}
\label{sec:results}

We first describe how participants engaged with \sysname{} across the fourteen-day deployment (Section~\ref{sec:overview}). We then examine emotion inference, momentary regulation, and cross-session personalization. RQ1 evaluates the accuracy of emotional states inferred from everyday narration. RQ2 examines emotional changes following guided conversations and subsequent guidance uptake. RQ3 examines how engagement and perceived personalization developed through continued use.

\subsection{Overview of \sysname{} Deployment and Engagement}
\label{sec:overview}

Participants completed 1{,}093 sessions and 10{,}953 conversation turns, averaging 57.5 sessions each ($SD = 11.8$, range 35--78), or 4.1 sessions per day. Just-in-time logs accounted for 55.0\% of sessions, morning check-ins 23.6\%, and evening reflections 21.4\%. Guided conversations averaged 10.02 turns ($SD = 2.80$), with a median session duration under three minutes. Participants used text in 872 sessions (79.8\%) and voice in 221 (20.2\%), with five participants using voice in more than 40\% of their sessions. Overall, participants rated the system as easy to use, with an exit rating of $M = 5.37$ ($SD = 1.54$) on the 7-point scale (Figure~\ref{fig:exit-rq3}).
Across sessions, self-reported pre-guidance valence was low ($\leq 4$) in 28.7\%, intermediate (5--6) in 33.0\%, and high ($\geq 7$) in 38.2\%. The router assigned 87.5\% of sessions to Cog and 12.5\% to Combined.

\begin{figure}[t]
\centering
\includegraphics[width=1.03\columnwidth]{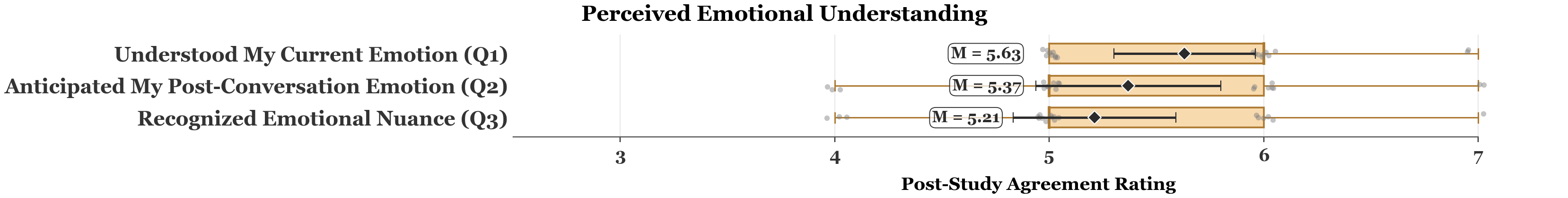}
\caption{Exit ratings of perceived emotional understanding. In-the-moment understanding item had the highest mean rating, while emotional nuance recognition had the lowest. Boxes show the distribution of participant ratings, grey points are individual participants, and diamonds with error bars show means with 95\% confidence intervals. Items were rated on a 7-point scale (1 = strongly disagree, 7 = strongly agree).}
\Description{Horizontal box plot with three rows, one per questionnaire item, on a one-to-seven scale. All three boxes sit between five and six. The in-the-moment item has the highest mean and the tightest spread.}
\label{fig:exit-rq1}
\end{figure}

\subsection{RQ1: Text-Based Appraisal Inference Supports Valence and Dominance but Remains Limited for Arousal}
\label{sec:rq1}

To address RQ1, we compared the system's estimates with participants' unanchored self-reports of the same moments. We assessed fine-grained accuracy across all paired observations (Table~\ref{tab:accuracy}), compared the appraisal scaffold with direct LLM inference and a lexicon baseline, and tested agreement after reducing estimates to lower-versus-higher states.

\begin{table}[t]
\caption{Estimation accuracy against unanchored self-report on the 9-point SAM scale. Initial estimates are evaluated against before-guidance ratings and post-guidance estimates against after-guidance ratings. The bottom three rows compare initial valence estimators across all narrations, with Claude Opus 4.8 used for both LLM conditions. Signed error is computed as estimate minus self-report, so positive values indicate overestimation. All correlations are significant at $p < .001$ (***).}
\label{tab:accuracy}
\small
\begin{tabular}{llrrr}
\toprule
& \textbf{Dimension / Estimator} & \textbf{MAE} & \textbf{$r$} & \textbf{Signed err.} \\
\midrule
\multirow{3}{*}{Initial}
& Valence   & 1.196 & .669*** & $+0.213$ \\
& Dominance & 1.299 & .538*** & $-0.061$ \\
& Arousal   & 1.651 & .236*** & $-0.129$ \\
\addlinespace
\multirow{3}{*}{Post-guidance}
& Valence   & 1.115 & .614*** & $-0.128$ \\
& Dominance & 1.189 & .495*** & $-0.393$ \\
& Arousal   & 1.963 & .175*** & $-1.097$ \\
\addlinespace
\midrule
\multirow{3}{*}{Valence estimators}
& NRC-VAD lexicon, 5-fold CV & 1.679 & .318*** & --- \\
& LLM without scaffold & 1.334 & .640*** & --- \\
& \sysname{} appraisal scaffold    & 1.196 & .669*** & --- \\
\bottomrule
\end{tabular}
\end{table}
\subsubsection{Appraisal-Guided Inference from Daily Narration Yielded Strong Valence and Moderate Dominance Correlations with Self-Reports}
\label{rq1valence}

Appraisal-guided inference from everyday narration showed a strong correlation with self-reported valence and a moderate correlation with self-reported dominance. Initial estimates correlated with self-reports at $r = .669$ for valence and $r = .538$ for dominance, with MAE of 1.196 and 1.299 points, respectively (Table~\ref{tab:accuracy}). Both dimensions showed lower estimation error than arousal, and valence remained the most accurately estimated dimension after guidance. Across all narrations, appraisal-guided inference achieved a valence correlation of $r = .669$ and an MAE of 1.196, compared with $r = .640$ and MAE = 1.334 for the same-model baseline without the appraisal scaffold and $r = .318$ and MAE = 1.679 for the cross-validated NRC-VAD baseline. Absolute valence error showed no statistically reliable trend across the 14-day deployment ($b = -0.0069$ on the log1p scale, $p_{\mathrm{FDR}} = .087$).

Participants also rated the companion highly for understanding how they felt in the moment ($M = 5.63$, $SD = 0.68$), anticipating how they would feel after conversations ($M = 5.37$, $SD = 0.90$), and recognizing emotional nuance ($M = 5.21$, $SD = 0.79$) on 7-point scales (Figure~\ref{fig:exit-rq1}). Their interview accounts showed that this understanding was most salient when \sysname{} inferred emotion from events containing little explicit emotion language. Fourteen of nineteen participants described cases in which \sysname{} inferred feelings or their causes without explicit emotion language. P1 said the system could \textit{``go straight from my description to dissecting what was going on inside me.''} P10 similarly explained that even when she did not have words for how she felt, the system could identify both the emotion and \textit{``what was responsible''} for it. Others described the system as helping \textit{``peel back those layers''} (P3), while P19 valued its recognition that \textit{``saying goodbye can still make you feel sad''} within an otherwise positive experience. Participants thus valued inference when it moved from an event description to an emotional interpretation, especially when it supplied a connection they had not already articulated.
By contrast, participants identified shallow restatement and miscalibrated emotional intensity as two recurring limitations in \sysname{}'s emotion inference. Intensity emerged as the more common boundary. In some cases, the system added little beyond explicit cues. P18 reported needing to state \textit{``I'm very sad''} before receiving a less flat interpretation, while P5 said presenting an obvious restatement as deeper insight \textit{``felt like it was provoking me.''} More often, the system identified the emotional direction but misjudged its magnitude. This distinction between emotional direction and intensity becomes especially visible in arousal inference, which we examine in Section~\ref{rq1arousal}.

\subsubsection{Text-Only Inference Yielded Weak Arousal Correlations and Recurrent Misjudgments of Emotional Intensity}
\label{rq1arousal}

Arousal estimates showed lower agreement with participants' self-reports across both fine-grained and binary evaluations, with mismatches concentrated in emotional intensity when expressive cues were unavailable in text. Although participants used voice in 221 sessions, \sysname{} transcribed speech before inference and did not retain prosodic cues. Its arousal estimates therefore reflected what participants said in the transcript without information from how they said it. Arousal estimates correlated weakly with self-report ($r = .236$; Table~\ref{tab:accuracy}). Binary estimates showed moderate agreement for valence ($\kappa = .539$) and dominance ($\kappa = .430$), but only slight agreement for arousal ($\kappa = .154$), and changing the cut point did not raise arousal agreement above $\kappa = .21$. The word-based baseline showed little correspondence with self-reported arousal ($r = .016$), compared with a higher correlation for valence ($r = .318$). 

Interview accounts identified recurrent amplification or flattening of emotional intensity. Eight of nineteen participants described cases in which \sysname{} identified the general emotional direction but misjudged how strongly they felt it. P12 said the system \textit{``definitely exaggerate[d] things that I wasn't feeling''} and described her emotions as being taken \textit{``way too extreme''}. Others distinguished wanting to express frustration from actually feeling intense distress, such as \textit{``I'm just trying to vent. I'm not actually mad''} (P3) and \textit{``Sometimes you're not really that upset over it, you just want to talk about it''} (P7). Participants also linked these mismatches to missing expressive cues, noting that without face-to-face information it was harder to judge \textit{``how upset you are''} (P7). These errors also changed how some participants communicated with the system. P12 softened her wording to avoid amplification, while P18 made emotional intensity more explicit when the initial interpretation felt too flat. The interview accounts thus located a recurring difficulty in judging emotional intensity from text. Because estimated intensity informed regulation-strategy selection, these reported mismatches were also relevant to the guidance examined in RQ2.

\begin{figure}[h]
\centering
\includegraphics[width=1.03\columnwidth]{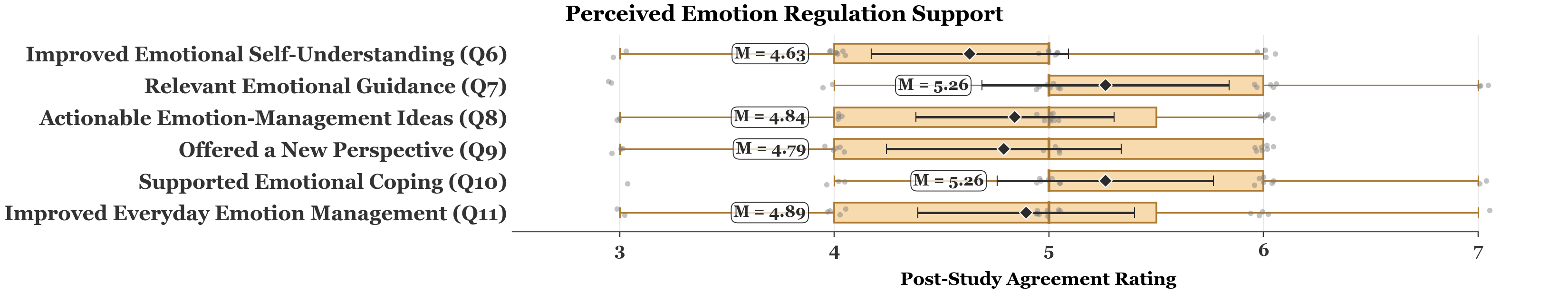}
\caption{Exit ratings of perceived emotion-regulation support. Relevant emotional guidance and support for emotional coping received the highest mean ratings (both $M = 5.26$), while improved emotional self-understanding received the lowest ($M = 4.63$). Boxes show participant rating distributions, grey points show individual participants, and diamonds with error bars show means with 95\% confidence intervals. Items were rated on a 7-point scale (1 = strongly disagree, 7 = strongly agree).}
\Description{A horizontal box plot with six emotion-regulation support items on a 1--7 self-rating scale. From top to bottom, the mean ratings are 4.63 for improved emotional self-understanding, 5.26 for relevant emotional guidance, 4.84 for actionable emotion-management ideas, 4.79 for a new perspective, 5.26 for emotional coping, and 4.89 for everyday emotion management. Grey points show individual participant ratings, boxes show their distributions, and black diamonds with error bars show means with 95\% confidence intervals.}
\label{fig:exit-rq2}
\end{figure}

\subsection{RQ2: Initial Emotional States, Conversational Responsiveness, and Practical Feasibility Shaped Momentary Regulation and Guidance Uptake}
\label{sec:rq2}

Guided conversations supported emotion regulation differently across participants' starting states and immediate needs. To address RQ2, we examine affective change by entry state, how exploration and emotional acknowledgment shaped perceived understanding, differences across regulation strategies, and when guidance was valued and adopted. We integrate pre- and post-guidance SAM ratings with strategy logs, session feedback, exit ratings, adoption probes, and interviews.

\subsubsection{State-Dependent Emotional Adjustment Following Guided Conversations Suggested Momentary Regulation}
\label{rq2repair}

\begin{figure*}[t]
\centering
\includegraphics[width=\textwidth]{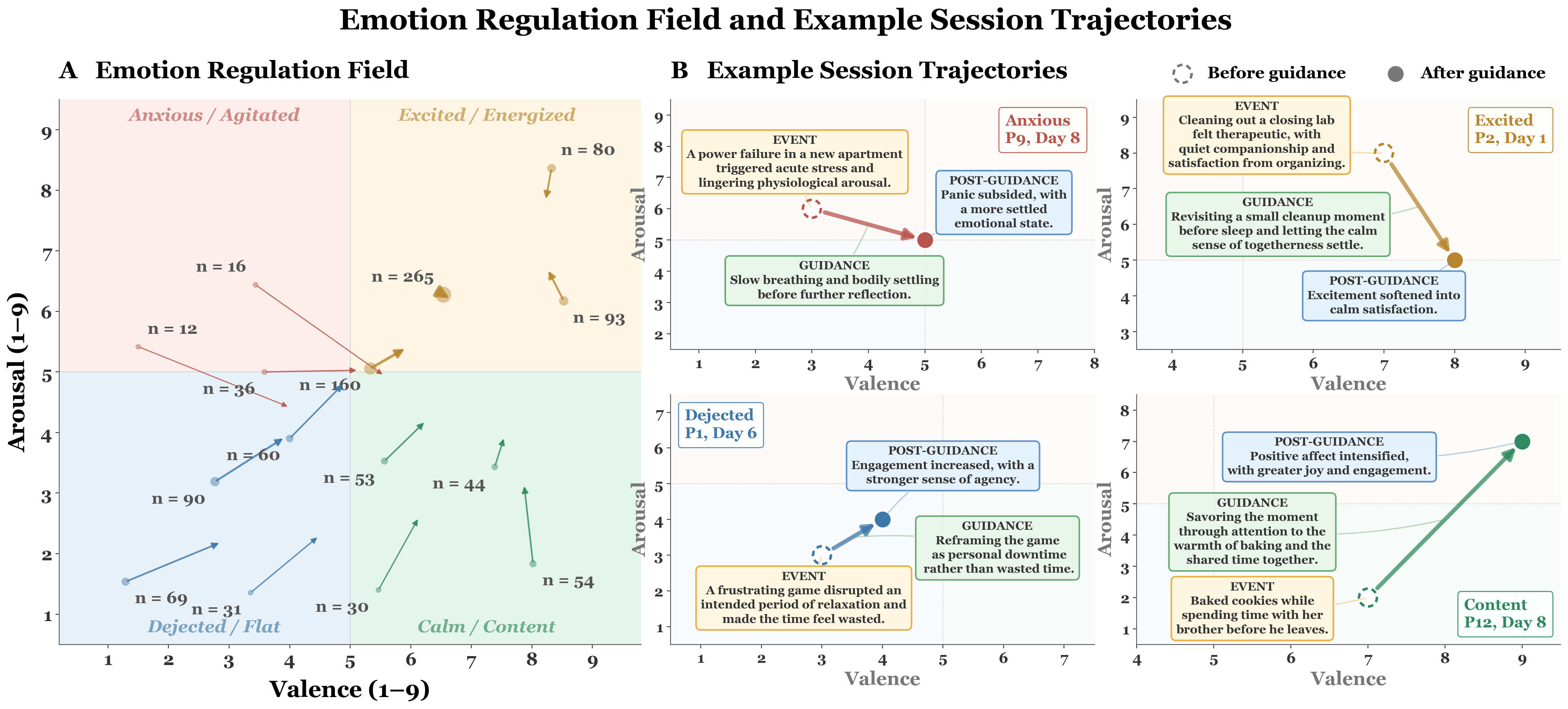}
\caption{Guided conversations were followed by state-dependent affective movement. (A) Arrows connect clustered mean self-reported valence and arousal before guidance to the corresponding means after guidance; labels show the number of sessions in each cluster. Anxious and dejected trajectories generally moved toward higher valence, with arousal decreasing for anxious states and increasing for dejected states. Calm trajectories also became more activated, while excited trajectories changed least. (B) Example session from each starting-state quadrant illustrates these directions by connecting the narrated event and guidance with the self-reported state after guidance.}
\Description{A two-part figure plotting valence horizontally and arousal vertically. Panel A divides the rating space into anxious or agitated, excited or energized, dejected or flat, and calm or content quadrants. Fifteen arrows connect the mean state before guidance, shown by a circle, to the mean state after guidance for clusters of nearby starting coordinates; each arrow is labelled with its session count. Most anxious arrows point right and downward, dejected and calm arrows point right and upward, and excited arrows are shorter and vary in direction. Panel B contains four example trajectories, one per quadrant. Each example shows a dashed circle for the state before guidance, a solid circle for the state after guidance, an arrow between them, and text boxes summarizing the event, guidance, and post-guidance experience.}
\label{fig:field}
\end{figure*}

Guided conversations were followed by state-dependent emotional adjustment consistent with momentary regulation, with the largest valence gains in anxious and dejected sessions and arousal shifting according to the starting state. Figure~\ref{fig:field} visualizes this pattern. Trajectories originating in the anxious quadrant generally moved toward higher valence and lower arousal, while those originating in the dejected quadrant moved toward higher valence and higher arousal. These shifts indicate different regulatory directions from different entry states, reducing activation when participants began anxious and increasing activation when they began dejected. Aggregated within each entry-state quadrant, anxious sessions gained $1.875$ points in valence ($SD = 1.453$) while decreasing $0.547$ points in arousal ($SD = 1.522$). Dejected sessions gained $1.208$ points in valence ($SD = 1.427$) and $0.780$ points in arousal ($SD = 1.220$). Valence gains were smaller in the two positive quadrants, at $+0.162$ in excited sessions and $+0.304$ in calm sessions, while arousal changed by $+0.069$ and $+0.906$, respectively.  The entry-valence split showed the same pattern. Sessions beginning at low valence ($\leq 4$) gained $1.344$ points ($SD = 1.455$), while those beginning at high valence ($\geq 7$) changed by $-0.108$ points ($SD = 0.857$). Across all sessions, valence increased by $0.525$, dominance by $0.469$, and arousal by $0.334$, with Wilcoxon tests at $p < .001$ (Appendix~\ref{app:marginal-affect}).  

Participants described the clearest benefit when conversations began from a state they wanted to change, while positive states were more often maintained. Exit ratings also reflected perceived regulation support, with participants reporting help in coping with their emotions ($M=5.26$, $SD=1.05$) and managing emotions in everyday life ($M=4.89$, $SD=1.05$; Figure~\ref{fig:exit-rq2}). Participants also reported support for understanding their own emotions ($M=4.63$, $SD=0.96$). P10 entered feeling \textit{``so drained''} and left much better, while P18 said the companion helped her reconsider a \textit{``really emotional''} situation and decide she did not need to be \textit{``this upset.''} P16 similarly found the conversation most helpful when she was \textit{``exhausted''} or needed someone to talk to in the moment. Positive or already-settled states changed less. P9 said such conversations helped \textit{``feel more positive afterward,''} while P12 \textit{``rarely ever mentioned a problem because I just didn't have any''} and P11 described her state as already normal and clear. These accounts complement the larger gains observed at low valence by distinguishing relief from distress from the maintenance of an already positive state.

\subsubsection{Contextual Exploration and Emotional Acknowledgment Before Advice Supported High Perceived Understanding}
\label{rq1understand}

Participants described feeling understood through contextual exploration and emotional acknowledgment before advice. Per-session understanding averaged 5.45 of 7, with 96.0\% of ratings at 5 or above, while correlations with absolute estimation error were small ($\rho = -.014$ for valence, $.029$ for arousal, and $.101$ for dominance). Because participants never saw the numerical estimates, their judgments reflected how understanding was conveyed through the interaction. The following findings describe how participants experienced understanding and how it related to emotional benefit. First, participants experienced understanding through an ECA sequence that elicited context before offering guidance. Fourteen participants described clarification and follow-up questions as helping the system reach an accurate interpretation. P3 said that \textit{``after going back and forth three, four times,''} the system had a good grasp of the situation. Logged ECA stages showed the same progression, with 99.9\% of sessions beginning in Exploration, an expected dwell time of 1.8 companion turns, and Comforting transitioning to Action in 72.2\% of adjacent stage pairs, compared with 16.1\% for direct Exploration-to-Action transitions. This sequence created space to clarify the event and respond to emotion before proposing action. When questions followed the wrong interpretation, however, the same structure could feel misaligned, as P8 described receiving questions that were not on \textit{``the right track.''}
Second, perceived understanding remained consistently high but did not increase reliably over time. A linear mixed-effects model estimated only a small day slope ($b = +0.0121$, $p_{\mathrm{FDR}} = .087$), equivalent to approximately $+0.16$ points from Day 1 to Day 14. Eighteen participants described the system as generally accurate, including P10, who felt understood \textit{``right from day one.''} Participants also linked this accuracy to disclosure. P2 evaluated the system relative to \textit{``how much information I gave it,''} while P5 sometimes needed to describe a situation in substantial detail before the system understood it. Stable understanding therefore emerged from an iterative exchange in which the system elicited relevant detail and participants supplied enough context for it to interpret.

\begin{figure*}[t]
\centering
\includegraphics[width=\textwidth]{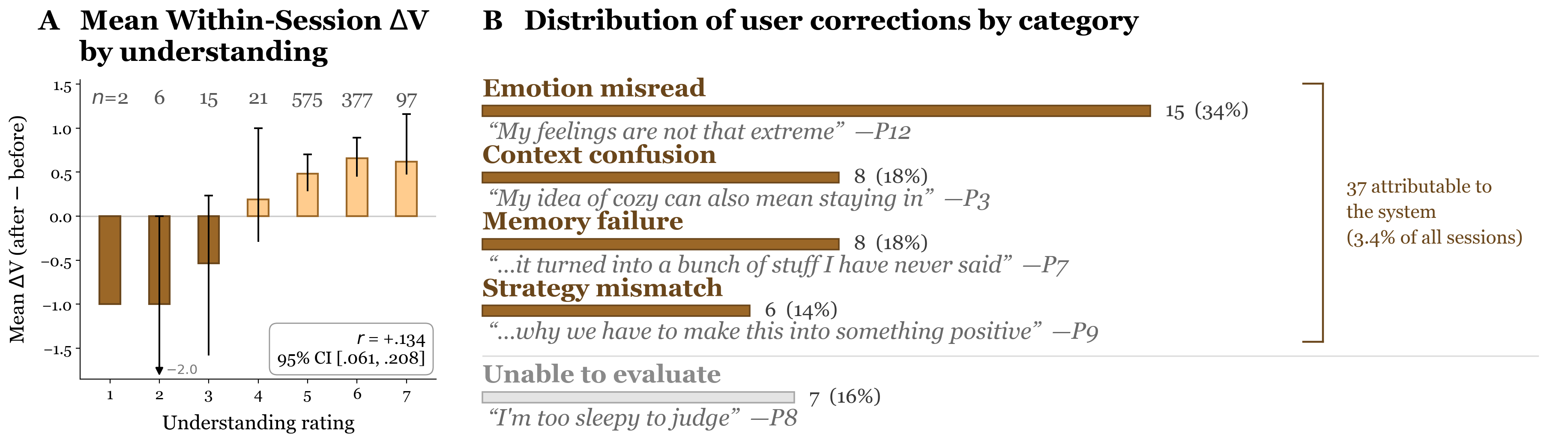}
\caption{Perceived understanding was associated with within-session valence change, and participant corrections identified recurring system mismatches. (A) Mean valence change (after minus before) by understanding rating. Bars show means, error bars show 95\% confidence intervals, and labels show session counts; higher ratings corresponded to greater valence gains ($r = .134$, 95\% CI $[.061, .208]$). (B) The 44 correction texts by coded category, each paired with a representative correction. The four system-attributable categories contained 37 corrections, representing 3.4\% of all sessions; the remaining seven indicated that the participant was unable to evaluate the system.}
\Description{A two-panel figure. Panel A is a bar chart of mean valence change after versus before guidance across understanding ratings from 1 to 7. Bars for ratings 1 to 3 extend below zero, while bars for ratings 4 to 7 extend above zero. Error bars show 95\% confidence intervals, and labels above the bars show 2, 6, 15, 21, 575, 377, and 97 sessions. Panel B is a horizontal bar chart of 44 correction texts grouped into emotion misread (15, 34\%), context confusion (8, 18\%), memory failure (8, 18\%), strategy mismatch (6, 14\%), and unable to evaluate (7, 16\%). Each category includes a representative participant correction. A bracket groups the first four categories as 37 corrections attributable to the system, or 3.4\% of all sessions.}
\label{fig:understanding}
\end{figure*}

Third, participants' corrections showed that perceived understanding extended beyond emotion labels to broader interactional alignment. Only 4.0\% of sessions included correction text, all following understanding ratings of 4 or below. Among 37 attributable corrections, 15 concerned emotion direction, intensity, or tone, 8 concerned the situation, 8 involved memory, and 6 concerned the desired support strategy (Figure~\ref{fig:understanding}B). P1 never felt a need to correct the system \textit{``on the part where it defines my emotions,''} whereas P2's corrections targeted \textit{``the wrong suggestion.''} Participants' sense of being understood therefore depended not only on emotion recognition, but also on contextual fit and appropriate support.
Fourth, perceived understanding was positively associated with within-session valence gain ($r = .134$, 95\% participant-bootstrap CI $[.061, .208]$). Mean valence change was negative at understanding ratings of 1--3 and positive at ratings of 5--7 ($+0.48$ to $+0.66$; Figure~\ref{fig:understanding}A). Sessions with correction text also showed lower valence change than uncorrected sessions ($-0.27$ vs. $+0.56$; paired Wilcoxon $W = 5$, $p = .005$, $n = 12$ participants), despite similar starting valence. Perceived misunderstanding was therefore associated with reduced emotional benefit.

\subsubsection{Contextual Strategy Selection Produced Distinct Regulatory Actions Without Detectable Differences in Immediate Outcomes}
\label{rq2grossmodelhelps}

Contextual strategy selection distributed guidance across the four Gross-informed strategy families, with no detectable differences in immediate valence gains (Friedman $\chi^2(3)=0.613$, permutation $p=.897$) or subjective evaluations across families. Mean within-session valence gain was highest for situation modification ($+0.672$, $n=317$), followed by cognitive change ($+0.585$, $n=246$), response modulation ($+0.453$, $n=256$), and attentional deployment ($+0.369$, $n=274$). Immediate evaluations were also similar across the four families, with mean helpfulness, feasibility, and willingness to act ratings ranging from 5.00 to 5.28 on the 7-point scales. Mixed-effects likelihood-ratio tests detected no significant differences by selected strategy (all $p>.29$, all $p_{\mathrm{FDR}}=.395$; Table~\ref{tab:strategy-feedback}). No strategy family showed a clear advantage in immediate outcomes under context-dependent selection.
Interviews confirmed that these strategy selections reached participants as different regulatory moves. Participants rated the companion's provision of actionable emotion-management ideas at $M=4.84$ ($SD=0.96$) and new perspectives at $M=4.79$ ($SD=1.13$; Figure~\ref{fig:exit-rq2}). For situation modification, P1 recalled being encouraged to start with the easiest part of a stalled task. Cognitive change used participants' own facts to revise an unfavorable interpretation. P19 described the companion countering self-blame by recalling what had already been completed that day, saying, \textit{``You already accomplished quite a lot.''} P12 adopted attentional suggestions such as getting a snack or taking a break. For response modulation, P16 drank warm water and breathed deeply before an exam, reporting that she \textit{``actually tried that, and it worked.''} The range of strategies did not ensure that every suggestion suited its context. Six participants described a fixed pool of suggestions, including P1, whose solitary leisure suggestion was applied to a relationship conflict where it \textit{``would only make things worse.''} These accounts show how the four families translated into concrete regulatory actions whose usefulness depended on their fit to the immediate situation.

\subsubsection{Practical Feasibility and Alignment with Immediate Support Needs Encouraged Greater Guidance Acceptance and Uptake}
\label{rq2adoption}

Participants carried guidance into subsequent action when it identified a concrete step that fit their immediate resources and constraints. Exit ratings of relevance averaged $5.26$ ($SD=1.19$; Figure~\ref{fig:exit-rq2}), while immediate feasibility and willingness-to-act ratings averaged $5.18$ ($SD=1.15$) and $5.13$ ($SD=1.23$), respectively (Table~\ref{tab:trends}). Evening probes documented selective uptake beyond the conversation. Among 44 categorized and manually checked responses, 24 described trying the guidance with a positive result and five described partial or alternative enactment, or mixed results. Four reported actions that coincided with what participants already intended to do, eight reported not trying the suggestion, and three could not recall it. Reported benefits included relief from a headache after taking a screen break and feeling better prepared after choosing an interview outfit.

Participants acted on suggestions that fit their available time and resources, while distinguishing a reasonable proposal from one they wanted to follow. P6, P16, and P1 described adopting suggestions they judged feasible or reasonable. P7 explained that \textit{``You can always take one minute to the side,''} while a walk could require time she did not have. The same practical criterion explained rejection. P3 could not obtain a suggested matcha at 5 a.m., and P12 would not interrupt a conversation with a friend to follow the advice. P18 articulated the distinction between willingness and reasonableness, saying, \textit{``Sometimes I did not really want to follow the suggestion, but I still thought the suggestion itself was reasonable.''}
Meanwhile, support was also judged by whether the moment called for reassurance, encouragement, or further listening. P2 sometimes wanted to hear that \textit{``it'll be okay,''} and reassurance was sufficient for that need. P18 instead described moments when she wanted the companion to \textit{``push me a little''} and help motivate action. Advice itself could be unwelcome when participants still wanted to talk. P7 explained that \textit{``sometimes you just want something to listen to you''} and that solutions could feel overwhelming. P11 likewise preferred the earlier conversation to the later advice. Useful guidance therefore required both a response suited to the current problem and recognition of the support the participant was ready to receive.

\subsection{RQ3: Conversational Familiarity Supported Engagement, While Cross-Session Memory Contributed to Personalization}
\label{sec:rq3}

To address RQ3, we examine changes in engagement and perceived helpfulness alongside participants' adjustments to their narration practices and support expectations. We then examine how remembered context, changing circumstances, and the persistence of corrections shaped perceived personalization. We combine longitudinal session logs and feedback with exit ratings and interviews across the 14-day deployment.

\begin{figure}[t]
\centering
\includegraphics[width=1.03\columnwidth]{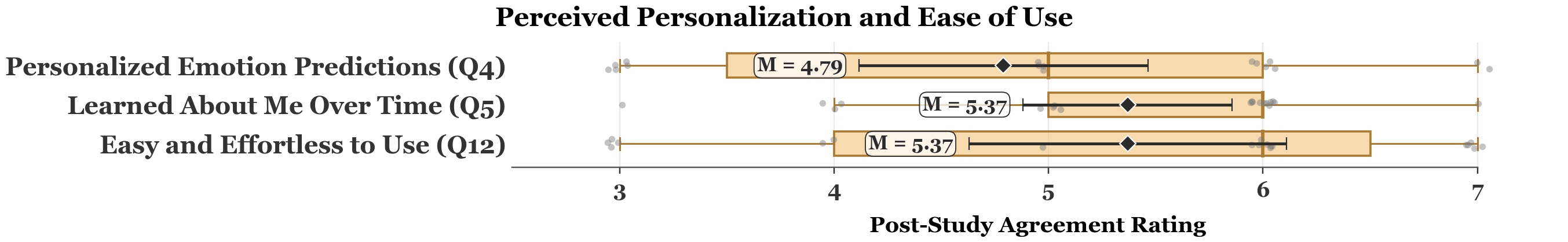}
\caption{Exit ratings of perceived personalization and ease of use over the 14-day deployment. Feeling increasingly known and ease of use received higher mean ratings than personalized emotion estimates. Items were rated on a 7-point scale (1 = strongly disagree, 7 = strongly agree).}
\Description{Horizontal box plot with three rows on a one-to-seven scale. Learning and ease of use have means near five and a half. Personalization has the lowest mean.}
\label{fig:exit-rq3}
\end{figure}

\subsubsection{Familiarity with Response Patterns Guided Narration Practices and Support Expectations}
\label{rq3engagement}

Across two weeks, participants continued to use \sysname{} at a similar frequency while writing less and rating its immediate helpfulness more highly. Daily session frequency showed no reliable trend ($p_{\mathrm{FDR}} = .963$), and the small decline in conversation turns did not remain significant after correction ($p_{\mathrm{FDR}} = .091$). Log-transformed mixed-effects models estimated daily decreases of 3.37\% in initial narration length and 4.30\% in user input length per turn, both measured in characters (both $p_{\mathrm{FDR}} < .001$). Helpfulness ratings increased over the two weeks ($b = +0.029$ points per day, $p_{\mathrm{FDR}} = .005$). Unadjusted means rose from 4.94 in Week 1 to 5.14 in Week 2 ($b = +0.216$ for the week contrast, $p_{\mathrm{FDR}} = .011$). Feasibility and willingness to act showed no significant day trends after FDR correction ($p_{\mathrm{FDR}} = .182$ and $.214$, respectively). Table~\ref{tab:trends} reports both models for all nine session measures.

Interviews revealed how participants adjusted the content, timing, and effort of their disclosure over time. Nine of nineteen participants described adapting their wording or conversational approach to how the companion responded. P5 spent increasing time composing initial narrations to \textit{``use as few words as possible''} while retaining relevant context. P15 similarly learned \textit{``how it understood messages over time''}, while P18 supplied additional detail and learned to \textit{``explicitly describe my current emotion''} to improve understanding. P8 deliberately shortened openings to invite follow-up questions, explaining that \textit{``I would know to wait''}, and left elaboration for subsequent turns. P14 described how the companion \textit{``finds a way to expand it''} after a short answer and follows existing details after a longer one. After receiving repetitive advice, P16 moved from detailed narration to sharing less, saying, \textit{``as time passed, I didn't share that much.''} These accounts distinguished deliberate adjustments to expression from reduced willingness to elaborate after repetitive advice.

Participants' expectations and evaluations of the companion also evolved over the two weeks. Three participants connected their evaluations to changing goals, familiarity, or their own interaction skills. P5 shifted from seeking empathy toward \textit{``practical suggestions''} and reported that approaching conversations with the goal of \textit{``looking for an answer or solution''} tended to leave him feeling better afterward. When explaining an emotional-understanding rating, P8 described \textit{``almost identical''} responses becoming more acceptable through \textit{``a basic kind of natural acceptance towards the AI''}. P12 could not distinguish whether the model had improved or \textit{``I interacted with it better''}. Participants thus evaluated later interactions through changing support goals, familiarity with the responses, and their own experience of communicating with the system.

\subsubsection{Relevant Recall Supported Personalization, While Outdated Context and Forgotten Corrections Disrupted Continuity}
\label{rq3personalization}

Participants judged personalization by whether the companion retained relevant personal context and used it appropriately in later interactions. Post-study ratings for the companion's ability to develop an understanding of the participant over time averaged $5.37$ ($SD = 1.01$), while ratings of personalized emotion estimates averaged $4.79$ ($SD = 1.40$; Figure~\ref{fig:exit-rq3}). Interviews reflected this variation. Ten of nineteen participants described personalized responses, while fourteen also encountered generic or scripted ones. Participants distinguished between simply remembering personal information and using it meaningfully in later responses. P14 treated recall of hobbies, a pet, and work preferences as evidence that \textit{``It has a memory.''} P1 acknowledged that the companion retained prior statements but described the resulting response as \textit{``a fairly linear answer,''} expecting it to connect remembered information to personal tendencies and likely reactions. Personalization therefore depended on how remembered context was interpreted and applied, not only on whether it was retained.

Continuity became most visible when the companion connected earlier experiences to later developments. P5 began feeling better understood \textit{``around halfway through the study,''} while P14 was surprised when the companion recalled \textit{``a small thing I'd mentioned only once''} one or two weeks earlier. P19 described a stronger form of continuity. After discussing a routine of going straight home and feeling inactive after work, P19 began visiting a caf\'e instead. The companion later recalled this change and recognized it as \textit{``a very good change.''} These cases demonstrated that remembered context supported personalization when it carried specific experiences forward and incorporated subsequent changes.

Personalization weakened when remembered context was missing, temporally misplaced, or treated as fixed. Twelve of nineteen participants reported instances in which the companion forgot prior context, and six described confusion about when events occurred or how they were related. P7 found that a later check-in no longer retained the context of a presentation discussed earlier that day. P10 similarly had to reconstruct information provided that morning, saying, \textit{``I have to list it all over again.''} P2 described an event from three days earlier being treated as current even after correction. Context could also persist too rigidly. P15 felt the companion \textit{``relied too heavily on what I said in the onboarding session,''} turning occasional activities into recurring assumptions despite \textit{``I don't do them like every day.''} These failures required participants to re-establish context or correct an outdated interpretation.
Moreover, participants also judged the incorporation of corrections by whether later responses reflected their feedback. Six participants reported that feedback was incorporated, while nine described corrections that did not persist or eventually stopped correcting the system. P12 noticed that feedback about exaggerated emotional intensity led to more frequent use of \textit{``calmer''} language. P18 also observed improvement, but only after having to \textit{``correct it two or three times.''} P4 described the opposite pattern, where the companion responded appropriately to a clarification in the moment but failed to retain it later. These experiences made the persistence of revised interpretations across encounters a distinct criterion for perceived personalization.

\section{Discussion}
\label{sec:discussion}

\subsection{Summary of Results}

Our fourteen-day deployment connects momentary emotion understanding and regulation with the cross-session demands of sustained support. For \textbf{RQ1 (Emotion Understanding)}, appraisal-guided estimates aligned more closely with self-reported valence and dominance than arousal. The scaffold also improved valence accuracy over both baselines. For \textbf{RQ2 (Regulation Guidance)}, guided conversations were followed by higher valence, lower arousal in anxious sessions, and higher arousal in dejected sessions. Participants felt understood through contextual exploration and emotional acknowledgment, and adopted guidance suited to their immediate needs and practical constraints. No strategy family showed a clear advantage in immediate outcomes under contextual selection. For \textbf{RQ3 (Cross-Session Continuity)}, inputs shortened and helpfulness ratings increased, while session frequency showed no reliable trend. Participants adjusted their narration and support expectations over time. Relevant recall and context updates reinforced perceived personalization, while outdated assumptions and forgotten corrections required users to repeat information or feedback. These findings motivate systems that adapt guidance to current needs and prior attempts while maintaining inspectable, revisable memory across changing circumstances.

\subsection{Study Limitations and Open Challenges for Sustained Emotional Support}
\label{sec:discussion-limitations}

\subsubsection{Study Limitations} 
\label{sec:discussion-study-limitations}
Our study has several limitations concerning the scope of the deployment and the strength of the resulting evidence. The fourteen-day study included 19 adults aged 18--33 who were generally familiar with AI, limiting the populations represented. The fixed check-in schedule also constrained naturalistic use and does not reflect how often the system would be used voluntarily. Beyond deployment scope, affect ratings rely on participants' own use of the scale, and concentrated understanding ratings reduce sensitivity to variation across sessions and participants. Although initial ratings were collected before system feedback, post-guidance reports may reflect how the conversation framed the experience~\cite{kim2024diarymate}. Comparisons across regulation strategies are similarly observational, since strategy selection depended on the participant's state. Controlled comparisons with unguided journaling and longer follow-up could strengthen these inferences and establish whether momentary effects carried into everyday well-being.

\subsubsection{Open Challenges for Sustained Emotional Support}
\label{sec:discussion-open-challenges}

Our findings identify reliable interpretation of emotion and interactional context, along with memory-based adaptation, as two open challenges for sustained emotional support. Text-based estimates showed lower agreement for arousal than for valence and dominance, while interpreting subtle or recurring emotions also requires context beyond the immediate event. Access to this context depends on what users choose to disclose. Privacy and judgment concerns can restrict disclosure~\cite{chiu2026privacy}, while the companion's questions and interpretations can shape how users describe subsequent experiences. This challenge extends beyond affect estimation to relational judgment. Models can generate fluent emotional context in supportive exchanges while still misjudging whether it fits the emotional moment~\cite{quan2025humor}. Reliable support therefore requires interpreting evidence shaped by both selective disclosure and the interaction itself. Multimodal cues such as speech prosody could supplement text-based inference, while brief confirmation could let users refine interpretations before they inform guidance. These mechanisms should preserve users' control over disclosure without making increasingly detailed accounts a prerequisite for support.
The second challenge is translating memory into guidance that responds to prior attempts and changing circumstances. Participants encountered suggestions they had already considered and corrections that did not persist across encounters. These patterns show that recalling personal information does not ensure that subsequent support adapts to it. Maintaining and revising temporal context remains difficult for long-context systems~\cite{wu2025longmemeval}. The challenge is determining which prior experiences remain relevant and how they should change the next recommendation.

\begin{figure}[t]
\centering
\includegraphics[width=\columnwidth]{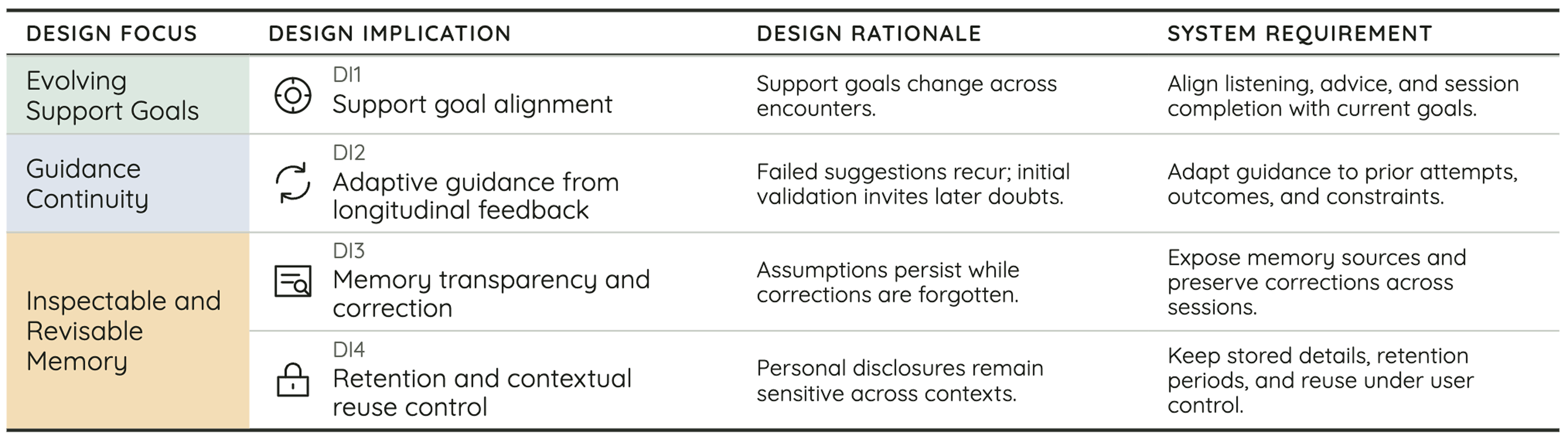}
\caption{Four design implications for sustained emotional support, organized around evolving support goals, guidance continuity, and user-controlled memory. Each implication connects its rationale to a system requirement.}
\Description{A four-column table presents design focuses, design implications, rationales, and system requirements. Evolving support goals motivate aligning listening, advice, and session completion with current needs. Guidance continuity motivates adapting recommendations to prior attempts, outcomes, and constraints. User-controlled memory motivates two implications: making memory sources visible and corrections persistent, and giving users control over stored details, retention periods, and contextual reuse.}
\label{fig:DI}
\end{figure}

\subsection{Toward Sustained Emotional Support}
\label{sec:discussion-ethics}

The challenges identified above highlight the capabilities required to sustain emotional support as users' goals and circumstances evolve. Momentary support addresses immediate emotional needs through acknowledgment and contextual guidance. Sustained support additionally requires tracking changes in support preferences, incorporating the outcomes of prior guidance, and updating personal context across interactions. Support preferences can shift between emotional expression and practical assistance, while recurring concerns require reassessing guidance in light of prior attempts and their outcomes. Cross-session personalization also raises questions of informational control, including how users inspect and correct stored information and govern its retention and reuse. Building on our findings, we derive several design implications for 1) aligning support with evolving goals, 2) adapting guidance through longitudinal feedback, and 3) supporting personalization and privacy through user-controlled memory (Figure ~\ref{fig:DI}).

\subsubsection{Changing Support Goals Require Flexible Transitions Between Listening and Guidance}

Sustained emotional support requires alignment between users' current goals and the form and timing of assistance. Our findings indicate that these goals vary across encounters and can change with continued use. P5 shifted from seeking empathy toward \textit{``practical suggestions,''} whereas P7 explained that \textit{``sometimes you just want something to listen to you''} and found solutions overwhelming. These accounts extend research on how users define chatbot support~\cite{song2024typing} by highlighting both changing preferences over time and the purpose of individual encounters. In \sysname{}, emotional intensity and controllability inform strategy selection, while dialogue progresses toward action within a turn budget. A strategy suited to the inferred emotion may still be premature when the user wants to continue expressing an experience. The Exploration--Comforting--Action framework permits flexible transitions~\cite{liu2021towards}, requiring systems to determine whether advice is appropriate at a given moment ~\cite{quan2026colleagues}. Yet LLMs can favor particular support strategies in ways that impair appropriate selection~\cite{kang2024preferencebias}. These findings motivate using the current support goal alongside emotional state to determine when to continue listening or introduce guidance.

\textbf{Design Implication 1: Support Goal Alignment through Intent Elicitation and Dialogue Adaptation.} Systems should identify current support goals from users' spontaneous requests or lightweight choices between continued expression and practical assistance. When intent remains unclear, targeted clarification questions can elicit conversational preferences~\cite{li2025gate}. These goals should remain revisable, with subsequent requests taking precedence over prior preferences. Before introducing advice, the system should confirm readiness if users have not already expressed it. Guidance should remain available on demand~\cite{song2025exploreself}, while declining advice should return the conversation to listening. Users should also be able to conclude the interaction once their needs are met. These controls align dialogue transitions and session completion with users' current goals across repeated interactions.

\subsubsection{Prior Attempts and Outcomes Should Inform the Next Step in Regulation Guidance}

Useful regulation guidance requires continuity between prior attempts, their outcomes, and subsequent recommendations. P18 reported receiving suggestions she had \textit{``already tried''} that \textit{``had not worked,''} while P3 valued familiar ideas made more salient through conversation. These accounts suggest that usefulness depends on how guidance builds on prior experience, including when to repeat, modify, or replace a suggestion. Regulatory flexibility theory similarly emphasizes responsiveness to feedback when maintaining or changing a strategy~\cite{bonanno2013flexibility}. Research on LLM support for procrastination further highlights the need for structured steps and adaptation to users' practical constraints~\cite{bhattacharjee2024procrastination}. Evaluating this adaptation also requires distinguishing immediate reassurance from subsequent usefulness. P8 described responses as \textit{``validating what I really wanted to hear''} but later questioned the behaviors they affirmed. These findings motivate preserving what users attempted and how they evaluated its consequences, so that later guidance can respond to both practical outcomes and retrospective judgments.

\textbf{Design Implication 2: Adaptive Guidance through Longitudinal Feedback on Attempts and Outcomes.} Systems should maintain a support history linking concerns and recommendations to attempted actions, reported outcomes, and barriers. Personalized guidance can draw on both historical behavioral data and contextual information elicited through dialogue~\cite{jorke2025gptcoach}. Brief follow-up could distinguish helpful attempts, unsuccessful attempts, and suggestions not yet tried. Consistent with adaptive interventions that use changing states and contexts to inform support decisions~\cite{nahumshani2018just}, these records could guide subsequent recommendations. Time constraints might prompt a shorter action, while an unsuccessful attempt could motivate an alternative or an adjustment to its implementation. Evaluation should track immediate emotional response, later uptake and outcomes, and whether subsequent guidance incorporates this feedback.

\subsubsection{Inspectable and Revisable Memory Should Give Users Control over Subsequent Personalization}

Sustained personalization depends on retaining personal context that remains accurate and that users are willing to retain and reuse. P15 found that the companion \textit{``relied too heavily on what I said in the onboarding session,''} treating occasional activities as recurring habits. P4 reported that a correction was understood immediately but \textit{``later it would forget again.''} These accounts highlight the need to revise both the interpretation of remembered information and its continued application. Research on conversational memory similarly documents users' concerns about retained information and their desire for greater visibility and control~\cite{chen2026memoryprivacy,zhang2025ragmemory}. Accuracy also intersects with privacy: users weigh the benefits of personalized responses against the risks of disclosing personal information~\cite{zhang2024fairgame}. Information shared to explain one experience may remain sensitive when recalled in another. Memory control therefore needs to address both whether the system's understanding remains valid and whether retaining and reusing that information remains acceptable to the user.

\textbf{Design Implication 3: Memory Transparency and User Control for Accurate Personalization.} Systems should expose memories as data objects that users can inspect, edit, and selectively reuse across conversations~\cite{huang2023memorysandbox}. Each entry should show its source conversation and relevant dates. Given that conversational memories can contain psychological inferences~\cite{dash2026selfportrait}, this view should distinguish users' explicit statements from system interpretations, alongside temporary states and recurring preferences. Users should be able to edit entries, mark events as completed, and remove inaccurate assumptions. Responses could link to the memories actually retrieved, making the basis for personalization traceable~\cite{zhang2025ragmemory}. Changes should propagate to related summaries and retrieval records so that subsequent guidance uses the revised context. Evaluation should check whether later conversations incorporate corrections and stop repeating outdated assumptions.

\textbf{Design Implication 4: Privacy Control over Memory Retention and Contextual Reuse.} Systems should let users choose session-only use or selective retention, review the details to be stored, and set retention periods. Memory selection should follow the support purpose~\cite{jo2024ltm}; unnecessary identifying details could be removed or generalized through user-reviewed suggestions~\cite{dou2024selfdisclosure}. Retained information should also have adjustable conditions for reuse. For example, a user could restrict a workplace disclosure to discussions of that concern while permitting selected scheduling details to inform sleep-related guidance. Findings that LLMs struggle with contextual privacy judgments~\cite{mireshghallah2024confaide} motivate checking both relevance and user permissions before retrieving a memory. Broader use should require permission, and expiration or deletion should remove the retained content from associated summaries and retrieval stores.

\subsection{Conclusion}
\label{sec}

Sustained emotional support requires generative agents to connect emotion understanding with guidance that adapts to users' changing needs and experiences. To evaluate these capabilities in everyday use, we deployed \sysname{} for fourteen days with 19 participants across 1{,}093 sessions. The companion integrated appraisal-guided emotion inference, regulation guidance, and cross-session memory. We paired system estimates with participants' self-reports and examined emotional changes alongside guidance uptake and experiences of continued use. Estimates corresponded more closely to self-reported valence and dominance than arousal. Perceived understanding showed little association with estimation error and instead emerged through contextual exploration and emotional acknowledgment. Guided conversations were followed by state-dependent emotional changes, while guidance uptake reflected immediate support needs and practical feasibility. Across repeated interactions, personalization relied on applying remembered experiences and user corrections as circumstances and expectations changed. These findings characterize sustained emotional support as a process in which agents continually interpret experiences, learn from prior interactions, and adjust their responses over time. This work offers a path toward affective intelligent agents that support everyday emotional life through effective momentary support and meaningful continuity over time.

%TC:ignore
\bibliographystyle{ACM-Reference-Format}
\bibliography{reference}

\appendix

\section{Runtime Orchestration}
\label{app:orchestration}

\begin{figure*}[h]
\centering
\includegraphics[width=\textwidth]{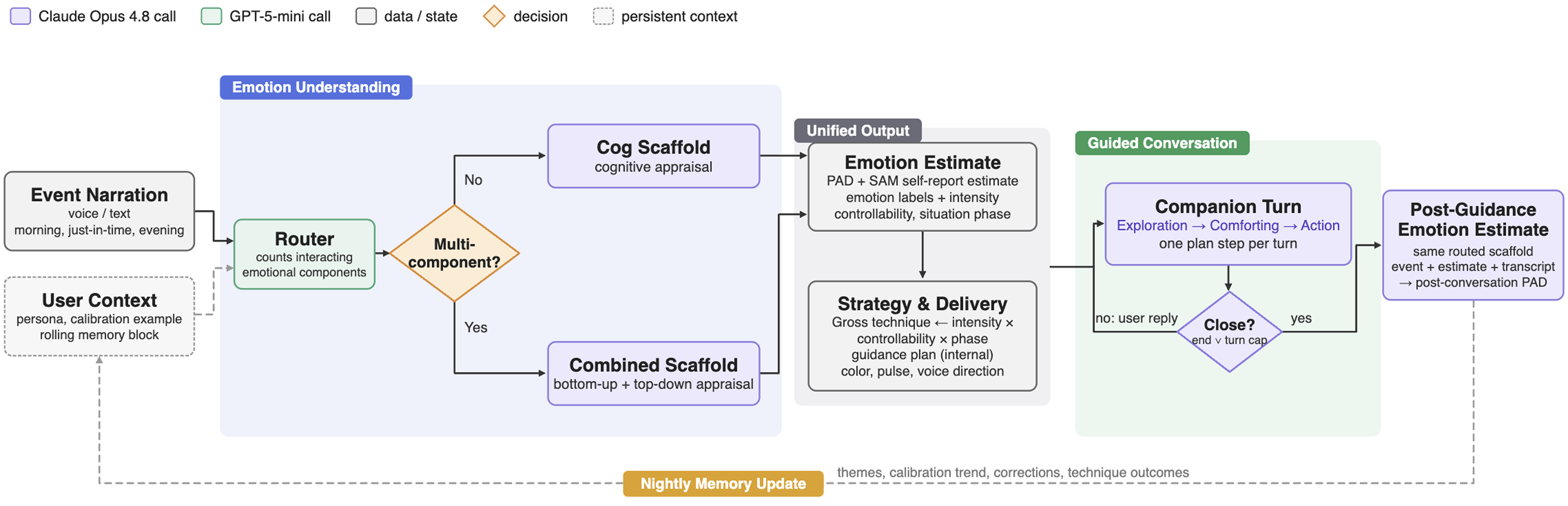}
\caption{Runtime orchestration for a single session. An event narration and the persisted user context enter a router that counts interacting emotional components and sends the narration to the cognitive-appraisal scaffold or, where several components interact, to the combined scaffold. Both produce one unified output: a dimensional emotion estimate with a matching SAM reading, emotion labels, appraised intensity, controllability, and situation phase, from which the regulation strategy and all delivery parameters are derived. The guided conversation then advances one plan step per turn along the exploration, comforting, and action arc until it closes. The scaffold selected by the router then re-runs over the event, the estimate, and the transcript to produce a post-guidance estimate. A nightly pass folds themes, calibration trend, corrections, and technique outcomes back into the user context.}
\Description{A left-to-right flow diagram of one session. Event narration and a persistent user context block feed a router, which branches on whether the narration contains multiple interacting emotional components and selects one of two appraisal scaffolds. Both scaffolds converge on a unified output box holding the emotion estimate and a strategy and delivery box. A guided conversation loop follows, cycling companion turns until a close condition is met, and its transcript feeds a post-guidance estimate. A dashed return path labelled nightly memory update runs from the end of the pipeline back to the user context.}
\label{fig:orchestration}
\end{figure*}

Figure~\ref{fig:orchestration} summarizes how the system routes event narrations, delivers guidance, estimates emotion after the conversation, and updates memory for subsequent sessions.

\section{Prompt Evaluation Details}
\label{app:prompteval}

Table~\ref{tab:prompteval} reports the full ranking distribution from the formative human evaluation (Section~\ref{sec:prompteval}). Cog placed in the top two in 72\% of rankings and last in only 7\%, and won 84\% of its direct matchups against Baseline; Combined earned the most first-place finishes after Cog (34 vs.\ 39 of 90) and won 78\% of its matchups against Baseline. Neuro lost 76\% of its direct matchups against Cog, and Baseline ranked last in 59\% of rankings.

For each scenario, we determined whether each rater ranked Cog or Combined higher and used majority vote across the three raters to identify the preferred prompt, following Section~\ref{sec:prompteval}. Preferences split 15:15 across the 30 scenarios. Cog was preferred in 14 of 17 single-component scenarios, and Combined in 12 of 13 multi-component scenarios. Selecting Cog for single-component events and Combined for multi-component events therefore matched the majority preference in 26 of these 30 formative scenarios (87\%; Figure~\ref{fig:promptcoding}). The four mismatches involved rater disagreement. Scenario-level overall preferences coincided with those based on reasoning fidelity in all 30 scenarios and outcome plausibility in 27.

\begin{table}[h]
\caption{Ranking distribution of the four emotion-inference prompts across 90 scenario-rater rankings (30 scenarios $\times$ 3 raters, overall-preference dimension). Each cell counts how often a prompt placed at that rank in the blind bracket tournament; lower mean rank is better. Results were consistent across the outcome-plausibility and reasoning-fidelity dimensions, which are omitted for brevity.}
\label{tab:prompteval}
\begin{tabular}{lccccc}
\toprule
\textbf{Prompt} & \textbf{\#1} & \textbf{\#2} & \textbf{\#3} & \textbf{\#4} & \textbf{Mean rank} \\
\midrule
Cog      & 39 & 26 & 19 & 6  & 1.91 \\
Combined & 34 & 31 & 8  & 17 & 2.09 \\
Neuro    & 13 & 12 & 51 & 14 & 2.73 \\
Baseline & 4  & 21 & 12 & 53 & 3.27 \\
\bottomrule
\end{tabular}
\end{table}

\section{Persona Interview Questions}
\label{app:persona}

The persona interview was administered conversationally by the companion during onboarding. Participants answered by voice, with the transcript editable as text, and each question included optional hint prompts. The seven questions, in order, with the construct each targets:

\begin{enumerate}
    \item \textit{(self-description)} ``Alright, no more checkboxes. Let's just talk for a bit. Want to give me a quick intro of yourself, so I can get to know you better?''
    \item \textit{(typical day)} ``I love that. Now, tell me about your life right now --- what does a typical day look like?''
    \item \textit{(coping style)} ``I can picture that. When things feel heavy or tiring, what usually helps you cope?''
    \item \textit{(support sources)} ``That sounds like something important for you. When life feels hard, who or what do you usually lean on?''
    \item \textit{(current preoccupations)} ``I'm glad to know what helps hold you up. What's been taking up a lot of space in your mind lately?''
    \item \textit{(good day)} ``I hear you. And on the other side --- what does a genuinely good day feel like for you?''
    \item \textit{(mood contagion)} ``Do other people's moods tend to rub off on you easily?''
\end{enumerate}

\section{Post-Study Questionnaire}
\label{app:questionnaire}

All items rated on a 7-point Likert scale (1 = strongly disagree, 7 = strongly agree).

\subsection*{Section 1: Perceived Emotional Understanding (RQ1)}
\begin{enumerate}
    \item The companion understood how I was feeling in the moment, based on what I described.
    \item The companion accurately anticipated how I would feel after our conversations.
    \item The companion picked up on the nuances of how I was feeling, not just the obvious emotions.
    \item The companion's emotion predictions felt personalized to me rather than generic.
    \item The companion seemed to get to know me better over the two weeks.
\end{enumerate}
Q1 and Q3 adapted from PETS \cite{schmidmaier2024pets}; Q2, Q4, and Q5 developed for this study.

\subsection*{Section 2: Perceived Emotional Guidance (RQ2)}
\begin{enumerate}
    \setcounter{enumi}{5}
    \item The companion helped me better understand my own emotions.
    \item The companion's guidance was relevant to my emotional situation.
    \item The companion's guidance gave me concrete ideas for managing my emotions.
    \item The companion's guidance helped me see my situation from a different perspective.
    \item The companion supported me in coping with emotional situations.
\end{enumerate}
Q7 adapted from the Session Rating Scale \cite{duncan2003session}; Q9 derived from the cognitive-reappraisal construct \cite{gross1998antecedent}; Q10 adapted from PETS \cite{schmidmaier2024pets}; Q6 and Q8 developed for this study.

\subsection*{Section 3: Technology Acceptance}
\begin{enumerate}
    \setcounter{enumi}{10}
    \item Using the companion helped me manage my emotions more effectively in everyday life.
    \item The companion was easy to use, and interacting with it did not require much effort.
\end{enumerate}
Q11 and Q12 adapted from the Technology Acceptance Model \cite{davis1989perceived}.

\section{Semi-Structured Interview Protocol}
\label{app:interview}

The interview lasted approximately 20--25 minutes and was audio-recorded. Participants first explained their ratings on the post-study questionnaire, then took part in an open-ended interview reflecting on their experiences and responding to the following questions.

\begin{enumerate}
    \item \textbf{Overall experience.} How would you describe your experience over the past two weeks?
    \item \textbf{Surprise moments.} Was there a specific moment when the companion's response surprised you? What made it right or wrong?
    \item \textbf{Emotion understanding.} How accurate did the system's emotion interpretations feel? Did accuracy change over time? Were the explanations meaningful and specific?
    \item \textbf{Suggestion adoption.} Which suggestions did you try? What worked, what didn't? Any suggestions that felt reasonable but you didn't act on?
    \item \textbf{Emotional awareness.} Did using the companion change how you notice or think about your emotions? Any new patterns you became aware of?
    \item \textbf{Most/least useful.} What was most useful? Least useful or most frustrating? \textit{Probes:} the colors, expressions, and movement; the burden of the daily check-ins; desired changes.
    \item \textbf{Open.} Anything else you'd like to share?
\end{enumerate}

\section{Participant Characteristics}
\label{app:participants}

Table~\ref{tab:participants} reports the demographic, AI-background, and baseline expressive-suppression measures collected in the onboarding questionnaires (Section~\ref{sec:participants}).

\begin{table}[h]
\caption{Participant characteristics from the onboarding questionnaires ($N = 19$). AI familiarity was rated on a 5-point scale from not familiar at all to extremely familiar. The three ERQ-S expressive-suppression items were rated on a 7-point scale from strongly disagree to strongly agree; the composite is their mean (Cronbach's $\alpha = .74$).}
\label{tab:participants}
\small
\begin{tabular}{@{}p{0.62\linewidth}r@{}}
\toprule
\textbf{Characteristic} & \textbf{$n$ (\%) or $M$ (\textit{SD})} \\
\midrule
\multicolumn{2}{@{}l}{\textit{Demographics}} \\
\quad Age in years & 25.1 (5.1) \\
\quad Age range & 18--33 \\
\quad Woman & 13 (68\%) \\
\quad Man & 6 (32\%) \\
\quad Asian & 15 (79\%) \\
\quad White & 2 (11\%) \\
\quad Black or African American & 2 (11\%) \\
\quad High school or equivalent & 5 (26\%) \\
\quad Some college & 1 (5\%) \\
\quad Bachelor's degree & 2 (11\%) \\
\quad Master's degree & 9 (47\%) \\
\quad Doctoral degree & 2 (11\%) \\
\addlinespace
\multicolumn{2}{@{}l}{\textit{AI background}} \\
\quad Familiarity with AI tools (1--5) & 3.95 (0.91) \\
\quad Very or extremely familiar & 13 (68\%) \\
\quad Uses AI tools sometimes & 4 (21\%) \\
\quad Uses AI tools often & 6 (32\%) \\
\quad Uses AI tools very often & 9 (47\%) \\
\quad Prior use of AI for emotional support & 8 (42\%) \\
\addlinespace
\multicolumn{2}{@{}l}{\textit{ERQ-S expressive suppression (1--7)}} \\
\quad I keep my emotions to myself & 4.53 (1.65) \\
\quad I control my emotions by not expressing them & 3.84 (1.61) \\
\quad When I am feeling negative emotions, I make sure not to express them & 3.79 (1.62) \\
\quad Composite & 4.05 (1.32) \\
\bottomrule
\end{tabular}
\end{table}

\section{Data Collected}
\label{app:data}

Table~\ref{tab:data} summarizes the data collected per session and study phase.

\begin{table}[h]
\caption{Data collected per session and per study phase.}
\label{tab:data}
\small
\begin{tabular}{p{0.27\linewidth}p{0.63\linewidth}}
\toprule
\textbf{Source} & \textbf{Data} \\
\midrule
Onboarding (Day 0) & Demographics; AI background; ERQ suppression; persona interview responses; calibration narration + SAM ratings \\
\addlinespace
Self-report (per session) & Before/after SAM ratings (valence, arousal, dominance); three feedback items; understanding rating + correction text if $\leq$4; rotating-question responses \\
\addlinespace
System logs (per session) & Narration text and modality; initial and post-guidance affective estimates (PAD, SAM, emotion labels, intensity); selected regulation strategy; per-turn conversational stage and support-strategy labels; full transcript with timestamps; delivery-channel state (color, pulse speed, expression sequence, speech on/off, language); prompt version and persona snapshot \\
\addlinespace
Engagement telemetry & App opens; just-in-time logging frequency and timing; missed sessions; voice vs.\ text ratio; slider interaction dynamics \\
\addlinespace
Post-study (Day 14+) & 12-item questionnaire; semi-structured interview recording \\
\bottomrule
\end{tabular}
\end{table}

\section{Within-Session Changes in Self-Reported Affect}
\label{app:marginal-affect}

\begin{figure*}[h]
\centering
\includegraphics[width=\textwidth]{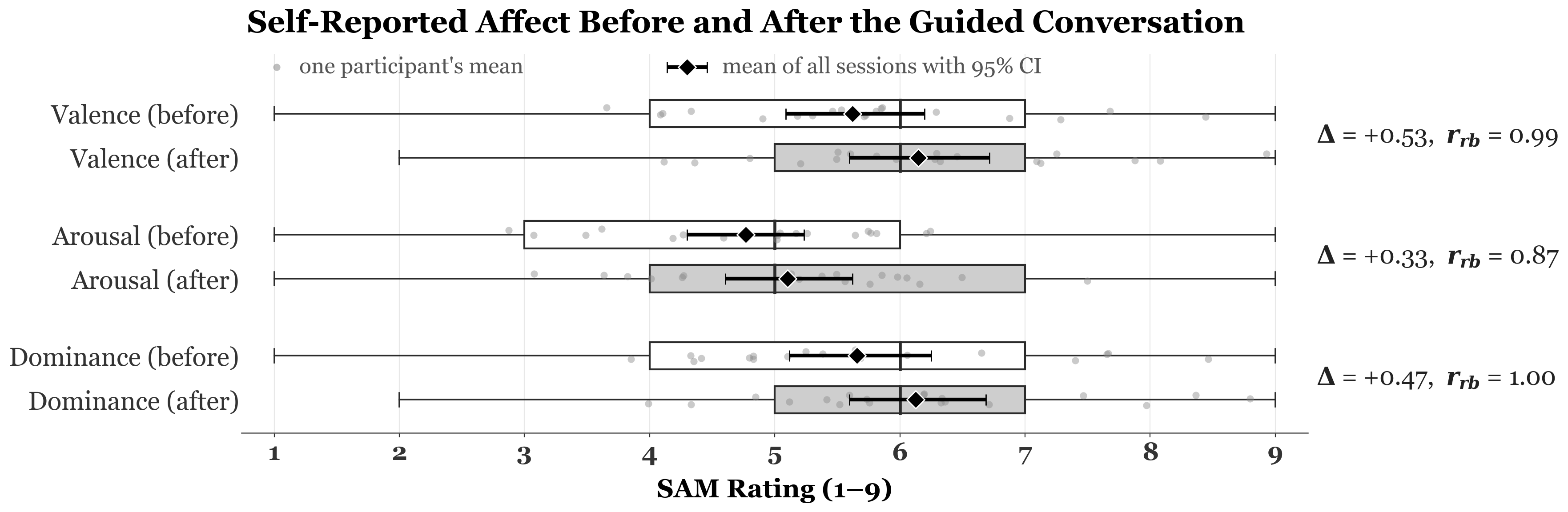}
\caption{Self-reported affect before and after guided conversation. Boxes show session-level distributions, grey points show each participant's mean, and diamonds with error bars show the session-weighted mean across sessions with 95\% participant-bootstrap confidence intervals. $\Delta$ denotes mean within-session change and $r_{\mathrm{rb}}$ the rank-biserial correlation from participant-level Wilcoxon signed-rank tests. All three contrasts remained significant after Holm correction ($p_{\mathrm{Holm}} < .001$).}
\Description{Six horizontal box plots on a 1--9 rating scale show valence, arousal, and dominance before and after guidance. For each dimension, the after-guidance distribution is shifted to the right of the before-guidance distribution. Grey points show participant means, and black diamonds with confidence intervals show session-level means. Labels report mean increases of 0.53 for valence, 0.33 for arousal, and 0.47 for dominance, with rank-biserial correlations of .99, .87, and 1.00, respectively.}
\label{fig:delta}
\end{figure*}

Figure~\ref{fig:delta} reports the marginal change in each SAM dimension, pooling sessions across the entry-state quadrants examined in Section~\ref{rq2repair}. Descriptive means are session-weighted; confidence intervals use 10{,}000 whole-participant bootstrap resamples. Inferential tests compare the 19 participants' mean changes against zero using exact two-sided Wilcoxon signed-rank tests, with Holm correction across the three dimensions.

For the regulation-field visualization (Figure~\ref{fig:field}), we clustered nearby pre-guidance coordinates within each entry-state quadrant and drew a vector from each cluster's mean starting state to its mean post-guidance state. We used three clusters for the smallest quadrant and four for each of the other quadrants to retain local variation without plotting every trajectory. One session from each quadrant illustrates a direction visible in the corresponding field and connects the reported event with the guidance and post-guidance state.

\section{Immediate Feedback by Regulation Strategy}
\label{app:strategy-feedback}

Table~\ref{tab:strategy-feedback} reports immediate feedback by selected strategy. For each rating, we fitted a linear mixed-effects model with strategy as a fixed effect and random intercepts for participant and participant-day. Likelihood-ratio tests compared models with and without the strategy term, and Benjamini--Hochberg correction controlled the false discovery rate across the three tests.

\begin{table*}[h]
\centering
\caption{Mean immediate feedback on 7-point scales by selected regulation strategy. SM denotes situation modification ($n=317$), CC cognitive change ($n=246$), RM response modulation ($n=256$), and AD attentional deployment ($n=274$). Means are unadjusted session averages. Test statistics refer to the strategy term in the mixed-effects models.}
\label{tab:strategy-feedback}
\small
\begin{tabular}{lrrrrrrr}
\toprule
Rating & SM & CC & RM & AD & $\chi^2(3)$ & $p$ & $p_{\mathrm{FDR}}$ \\
\midrule
Helpfulness & 5.03 & 5.16 & 5.01 & 5.00 & 3.00 & .391 & .395 \\
Feasibility & 5.28 & 5.18 & 5.20 & 5.04 & 2.98 & .395 & .395 \\
Willingness & 5.24 & 5.09 & 5.09 & 5.10 & 3.72 & .293 & .395 \\
\bottomrule
\end{tabular}
\end{table*}

\section{Session Measures Over the Deployment}
\label{app:trends}

Table~\ref{tab:trends} reports the full longitudinal results for the nine session measures summarized in Sections~\ref{sec:rq1} and \ref{sec:rq3}.
Each measure was modelled twice with the same random-effects structure, random intercepts for participant and for participant-day. The primary model treats study day as a continuous predictor and estimates a per-day slope. The secondary model replaces day with a week factor (W1 = days 1--7, W2 = days 8--14) and estimates the W2 minus W1 contrast. Sessions per day is a property of the participant-day and was modelled on daily aggregates; the other eight measures were modelled at the session level. Counts used a negative binomial family with a log link, narration length and characters per turn a Gaussian family on the log scale, and absolute valence error a Gaussian family on the log1p scale, so their estimates are on the link scale and correspond to $-0.50\%$, $-3.37\%$, $-4.30\%$ per day for conversation turns, narration length, and characters per turn. For log-linked outcomes, the table reports $100b$ on the link scale, while percentage changes in the main text are obtained as $100(\exp(b) - 1)$. The remaining measures, all 7-point ratings, used a Gaussian family on their observed scale. Benjamini--Hochberg correction controlled the false discovery rate across the nine tests within each model.
Separate exploratory models added participant-specific random slopes for study day to assess individual variation in change.

\begin{table*}[ht]
\centering
\caption{Longitudinal models for the nine session measures across the 14-day deployment. $M$ (SD) are unadjusted means over all sessions and within Week 1 (days 1--7) and Week 2 (days 8--14). $b_{\mathrm{day}}$ is the per-day slope from the day-continuous mixed-effects model (reported $\times 100$), and $\Delta_{\mathrm{week}}$ is the Week 2 minus Week 1 contrast from the week-binned model. Estimates for count and log-transformed outcomes are on the link scale.}
\label{tab:trends}
\footnotesize
\setlength{\tabcolsep}{4pt}
\begin{tabular}{@{}lrrrcrcr@{}}
\toprule
& \multicolumn{3}{c}{$M$ (SD)} & \multicolumn{2}{c}{Day-continuous model} & \multicolumn{2}{c}{Week-binned model} \\
\cmidrule(lr){2-4}\cmidrule(lr){5-6}\cmidrule(lr){7-8}
Measure & Overall & Week 1 & Week 2 & $b_{\mathrm{day}}\!\times\!100$ [95\% CI] & $p_{\mathrm{FDR}}$ & $\Delta_{\mathrm{week}}$ [95\% CI] & $p_{\mathrm{FDR}}$ \\
\midrule
\multicolumn{8}{@{}l}{\textit{Counts (negative binomial)}} \\
\quad Sessions per day    & 4.13 (1.06)  & 4.16 (1.13)  & 4.10 (1.00)  & $-0.04$ [$-1.52$, $1.45$]  & .963    & $-0.019$ [$-0.137$, $0.098$]  & .747 \\
\quad Conversation turns  & 10.02 (2.80) & 10.23 (2.93) & 9.87 (2.66)  & $-0.50$ [$-0.98$, $-0.02$] & .091    & $-0.037$ [$-0.075$, $0.001$]  & .109 \\
\addlinespace
\multicolumn{8}{@{}l}{\textit{Log-transformed (lengths in characters, $|e_V|$ in SAM points with log1p)}} \\
\quad Narration length    & 59.5 (105.9) & 69.9 (140.7) & 48.6 (45.0)  & $-3.43$ [$-4.75$, $-2.10$] & $<.001$ & $-0.234$ [$-0.341$, $-0.127$] & $<.001$ \\
\quad Characters per turn & 33.3 (32.0)  & 38.6 (37.0)  & 27.7 (24.6)  & $-4.39$ [$-5.48$, $-3.30$] & $<.001$ & $-0.284$ [$-0.375$, $-0.194$] & $<.001$ \\
\quad Valence error $|e_V|$ & 1.21 (1.08) & 1.27 (1.08) & 1.15 (1.07)  & $-0.69$ [$-1.42$, $0.03$]  & .087    & $-0.055$ [$-0.113$, $0.002$]  & .109 \\
\addlinespace
\multicolumn{8}{@{}l}{\textit{Ratings (1--7)}} \\
\quad Understanding       & 5.45 (0.81)   & 5.41 (0.77)   & 5.49 (0.85)   & $+1.21$ [$-0.04$, $2.45$]  & .087    & $+0.068$ [$-0.031$, $0.168$]  & .227 \\
\quad Helpfulness         & 5.04 (1.04)  & 4.94 (1.07)  & 5.14 (1.00)  & $+2.94$ [$1.12$, $4.76$]   & .005    & $+0.216$ [$0.070$, $0.361$]   & .011 \\
\quad Feasibility         & 5.18 (1.15)  & 5.12 (1.17)  & 5.23 (1.12)  & $+1.34$ [$-0.45$, $3.13$]  & .182    & $+0.110$ [$-0.033$, $0.252$]  & .196 \\
\quad Willingness to try  & 5.13 (1.23)  & 5.11 (1.27)  & 5.16 (1.19)  & $+1.23$ [$-0.61$, $3.06$]  & .214    & $+0.076$ [$-0.070$, $0.222$]  & .348 \\
\bottomrule
\end{tabular}
\end{table*}
%TC:endignore

\end{document}